 \documentclass[structabstract]{aa}  
\usepackage{graphicx}
\usepackage{txfonts}
\usepackage{natbib}
\bibpunct{(}{)}{;}{a}{}{,}

\begin{document}

   \title{The orbits of subdwarf-B + main-sequence binaries}

   \subtitle{II. Three eccentric systems; BD\,$+$29$^{\circ}$3070, BD\,$+$34$^{\circ}$1543 and Feige\,87}

   \author{J.~Vos
          \inst{1}
          \and
	  R.H.~\O{}stensen
	  \inst{1}
	  \and
	  P. N\'{e}meth
	  \inst{1}
	  \and
	  E.M.~Green
	  \inst{2}
	  \and
	  U.~Heber
	  \inst{3}
	  \and
	  H.~Van~Winckel
	  \inst{1}
          }


   \institute{
    Instituut voor Sterrenkunde, KU Leuven, Celestijnenlaan 200D, B-3001 Leuven, Belgium\\
    \email{jorisv@ster.kuleuven.be}
    \and
    Steward Observatory, University of Arizona, 933 North Cherry Avenue, Tucson, AZ 85721, USA
    \and
    Dr. Remeis-Sternwarte Astronomisches Institut, Universit\"{a}t Erlangen-N\"{u}rnberg, D-96049 Bamberg, Germany
    }

   \date{Received \today; accepted ???}
 
  \abstract
   {The predicted orbital-period distribution of the subdwarf-B (sdB) population is bi-modal with a peak at short ( $<$ 10 days) and long ( $>$ 250 days) periods. Observationally,  many short-period sdB systems are known, but the predicted long period peak is missing as orbits have only been determined for a few long-period systems. As these predictions are based on poorly understood binary-interaction processes, it is of prime importance to confront the predictions with reliable observational data. We therefore initiated a monitoring program to find and characterize long-period sdB stars. }
   {In this paper we aim to determine the orbital parameters of the three long-period sdB+MS binaries BD+29$^{\circ}$3070, BD+34$^{\circ}$1543 and Feige\,87, to constrain their absolute dimensions and the physical parameters of the components.}
   {High-resolution spectroscopic time series were obtained with HERMES at the Mercator telescope on La Palma, and analyzed to determine the radial velocities of both the sdB and MS components. Photometry from the literature was used to construct the spectral-energy distribution (SED) of the binaries. Atmosphere models were used to fit these SEDs and to determine the surface gravities and temperatures of both components of all systems. Spectral analysis was used to check the results of the SEDs.}
   {An orbital period of 1283 $\pm$ 63 d, a mass ratio of $q$ = 0.39 $\pm$ 0.04 and a significant non-zero eccentricity of $e$ = 0.15 $\pm$ 0.01 were found for BD+29$^{\circ}$3070. For BD+34$^{\circ}$1543 we determined $P_{\rm{orb}}$ = 972 $\pm$ 2 d, $q$ = 0.57 $\pm$ 0.01 and again a clear non-zero eccentricity of $e$ = 0.16 $\pm$ 0.01. Last, for Feige\,87 we found $P_{\rm{orb}}$ = 936 $\pm$ 2 d, $q$ = 0.55 $\pm$ 0.01 and $e$ = 0.11 $\pm$ 0.01.}
   {BD+29$^{\circ}$3070, BD+34$^{\circ}$1543 and Feige\,87 are long period sdB + MS binaries on clearly eccentric orbits. These results are in conflict with the predictions of stable Roche-lobe overflow models.}

   \keywords{stars: evolution -- stars: fundamental parameters -- stars: subdwarfs -- stars: binaries: spectroscopic}

   \maketitle
%

\defcitealias{Vos12}{Paper I}

\section{Introduction}\label{s-intro}
Hot subdwarf-B (sdB) stars are core helium burning stars with a very thin hydrogen envelope (M$_{\rm{H}}$ $<$ 0.02 $M_{\odot}$), and a mass close to the core helium flash mass $\sim$ 0.47 $M_{\odot}$ \citep{Saffer94, Brassard01}. These hot subdwarfs are found in all galactic populations, and they are the main source for the UV-upturn in early-type galaxies \citep{Green86,Greggio90, Brown97}. Furthermore, their photospheric chemical composition is governed by diffusion processes causing strong He-depletion and other chemical peculiarities \citep{Heber98}. 
The formation of these extreme horizontal branch objects is still puzzling. To form an sdB star, its progenitor needs to lose its hydrogen envelope almost completely before reaching the tip of the red giant branch (RGB), so that the core ignites while the remaining hydrogen envelope is not massive enough to sustain hydrogen shell burning. A variety of possible formation channels have been proposed. The earlier suggestions were based on single star evolution, e.g. stellar wind mass loss near the tip of the RGB \citep{Dcruz96} or enhanced mass loss due to rotationally driven helium mixing \citep{Sweigart97}.
It was found, however, that many sdB stars reside in binary systems \citep{Maxted01}, indicating that binary interaction plays an important role \citep{Mengel76}. Currently, there is a consensus that sdB stars are formed by binary evolution only, and several evolutionary channels have been proposed, where binary-interaction physics plays a major role. Close binary systems can be formed in a common envelope (CE) ejection channel \citep{Paczynski76}, while stable Roche-lobe overflow (RLOF) can produce wide sdB binaries \citep{Han00, Han02}. An alternative formation channel forming a single sdB star is the double white dwarf (WD) merger, where a pair of white dwarfs spiral in to form a single sdB star \citep{Webbink84}.

\citet{Han02, Han03} addressed these three binary formation mechanisms, and performed binary population synthesis (BPS) studies for two kinds of CE ejection channels, two possible stable RLOF channels and the WD merger channel. The CE ejection channels produce close binaries with periods of $P_{\rm{orb}}$ = 0.1 -- 10 d, and main-sequence (MS) or white-dwarf (WD) companions. The sdB  binaries formed through stable RLOF have orbital periods ranging from 10 to 500 days, and MS companions. An alternative stable RLOF channel based on the $\gamma$-formalism is described by \citet{Nelemans00, Nelemans01, Nelemans10} and can produce sdB binaries with periods on the order of $1-2$ years.  Finally, The WD merger channel can lead to sdB stars with a higher mass, up to 0.65 $M_{\odot}$. A detailed review of hot subdwarf stars is given by \citet{Heber09}.

Many observational studies have focused on short-period sdB binaries \citep{Koen98,Maxted00,Maxted01,Heber02,Morales03,Napiwotzki04,Copperwheat11}, and over 100 of these systems are currently known \citep[Appendix A]{Geier11}. These observed short-period sdB binaries correspond very well with the results of BPS studies. However, only a few long period sdB binaries are known \citep{Green01,Oestensen11,Oestensen12,Deca12,Barlow12,Vos12}, and the current studies show that there are still large discrepancies between theory and observations \citep{Geier13}. In a recent response to these discoveries \citet{Chen13} have revisited the RLOF models of \citet{Han03} with more sophisticated treatment of angular momentum loss. Their revised models show mass -- orbital period relations that increase substantially as a function of composition, with solar metallicity models reaching periods up to 1100\ d. They also note that by allowing the transfer of material extending beyond the classic Roche lobe (atmospheric 
RLOF) they can reach periods as long as $\sim$1600\ d.

In this paper we present the orbital and atmospheric parameters of the three long-period sdB + MS binaries BD+29$^{\circ}$3070, BD+34$^{\circ}$1543 and Feige\,87, using the methods described in \citet{Vos12}, hereafter Paper I. In Sect. \ref{s-spectroscopy} the radial velocities are determined for both components after which the orbital parameters are derived. Using the obtained mass ratio, the atmospheric parameters are derived from the spectral-energy distribution (Sect. \ref{s-SED}), and spectral analysis (Sect. \ref{s-specanal}). Furthermore, the surface gravity of the sdB component is estimated based on the gravitational redshift in Sect. \ref{s-gr}. Finally in Sect. \ref{s-absolutepar} and \ref{s-summary} all results are summarized.
BD+29$^{\circ}$3070, BD+34$^{\circ}$1543 and Feige\,87 are part of a long-term spectroscopic monitoring program, and preliminary results of these and five more systems in this program were presented in \citet{Oestensen11, Oestensen12}.

\section{Spectroscopy}\label{s-spectroscopy}
High resolution spectroscopic observations of BD+29$^{\circ}$3070, BD+34$^{\circ}$1543 and Feige\,87 were obtained with the HERMES spectrograph (High Efficiency and Resolution Mercator Echelle Spectrograph, R = 85\,000, 55 orders, 3770-9000 \AA, \citealt{Raskin11}) attached to the 1.2-m Mercator telescope at the Roque de los Muchachos Observatory, La Palma. HERMES is connected to the Mercator telescope by an optical fiber, and is located in a temperature controlled enclosure to ensure optimal wavelength stability. In \citetalias[Sect.\,2]{Vos12} the wavelength stability was checked, using 38 radial velocity standard stars of the IAU observed over a time span of 1481 days, and a standard deviation of 80 m~s$^{-1}$ with a non-significant shift to the IAU radial velocity standard scale was found. In total there were 31 spectra of BD+29$^{\circ}$3070, 30 of BD+34$^{\circ}$1543 and 33 of Feige\,87 taken between June 2009 and January 2013. 
The observations are summarized in Table \ref{tb-observations}. HERMES was used in high-resolution mode, and Th-Ar-Ne exposures were made at the beginning and end of the night. The exposure time of the science observations was adapted to reach a signal-to-noise ratio (S/N) of 25 in the $V$--band, whenever observing conditions permitted. The HERMES pipeline v5.0 was used for the basic reduction of the spectra, including barycentric correction.

For BD+34$^{\circ}$1543 there was one more high resolution spectrum available, taken with the FOCES spectrograph (Fiber-Optics Cassegrain Echelle Spectograph, R = 30\,000, 3600-6900 \AA) attached to the 2.2-m telescope at Calar Alto observatory, Spain. This spectrum was obtained in February 2000 (HJD = 2451576.5166). The spectrum was reduced as described in \citet{Pfeiffer98} using the IDL macros developed by the Munich Group.

Flux-calibrated spectra of BD+29$^{\circ}$3070, BD+34$^{\circ}$1543, and Feige\,87 were taken with the Boller and Chivens (B\&C) spectrograph attached to the University of Arizona's 2.3\,m Bok telescope located on Kitt Peak. All three stars were observed using a 2.5'' slit and 1st order 400/mm grating blazed at 4889 A, with a UV-36 filter to block 2$^{\rm{nd}}$ order light. These parameters provided a 9 \AA\ resolution over the wavelength range 3600-6900 \AA. BD+29$^{\circ}$3070 was observed on 25-06-2000 with an exposure time of 30 s, resulting in an overall S/N of 235 per resolution element (134 per pixel, and slightly higher in the range 3600-5000 \AA). BD+34$^{\circ}$1543 was observed once on 17-09-1998 and five additional times between 2005 and 2007, for a total exposure time of 260 s and a (formal) S/N $\sim$ 750 (435 per pixel). Feige\,87 was observed twice, on 10-03-1999 and 06-06-1999, for a total exposure time of 210 s and S/N of 345 (195 per pixel). The spectra were bias-subtracted, flat-fielded, 
optimally extracted, and wavelength calibrated using standard IRAF\footnote{IRAF is distributed by the National Optical Astronomy Observatory; see http://iraf.noao.edu/} tasks. 
They were flux calibrated using either BD+28$^{\circ}$4211 or Feige\,34 as flux standards. The individual spectra for BD+34$^{\circ}$1543 and Feige\,87 were combined by determining the cross-correlation velocities using only the Balmer and helium lines, and shifting each spectrum to the mean sdB velocity before combining (although the velocity shifts were always small, less than 1/3 of a pixel, compared to the spectral resolution 3.15 pixels). While these spectra have a too low resolution to obtain radial velocities, they are used to determine spectroscopic parameters of both components in Section \ref{s-specfit}.

\begin{table}
\caption{The observing dates (mid-time of exposure) and exposure times of the spectra of BD+29$^{\circ}$3070, BD+34$^{\circ}$1543 and Feige\,87}
\label{tb-observations}
\centering
\begin{tabular}{llllll}
\hline\hline
\multicolumn{2}{c}{BD+29$^{\circ}$3070} & \multicolumn{2}{c}{BD+34$^{\circ}$1543} & \multicolumn{2}{c}{Feige\,87}\\
BJD		&	Exp.	&	BJD		&	Exp.	&	BJD		&	Exp.	\\
--2450000	&	s	&	--2450000	&	s	&	--2450000	&	s	\\\hline
\noalign{\smallskip}
5029.4643	&	1800	&	5166.5252	&	700	&	5028.4488	&	1800	\\
5040.4568	&	1800	&	5218.5730	&	600	&	5028.4703	&	1800	\\
5055.4675	&	1200	&	5238.5505	&	900	&	5028.4917	&	1800	\\
5060.3980	&	1210	&	5238.5664	&	900	&	5251.7145	&	890	\\
5321.6137	&	1500	&	5238.5775	&	900	&	5251.7247	&	390	\\
5351.4739	&	1500	&	5240.5316	&	1800	&	5341.5722	&	1603	\\
5425.4024	&	1600	&	5475.6918	&	900	&	5351.4184	&	2700	\\
5619.7185	&	1500	&	5502.7281	&	900	&	5371.4175	&	2700	\\
5638.6611	&	1700	&	5507.6955	&	900	&	5566.7810	&	2700	\\
5648.6732	&	1500	&	5553.5767	&	900	&	5616.6402	&	2100	\\
5658.6816	&	1600	&	5614.5668	&	900	&	5622.7074	&	720	\\
5670.5225	&	1500	&	5621.5560	&	800	&	5622.7317	&	2000	\\
5678.5495	&	1940	&	5658.3476	&	594	&	5656.6032	&	911	\\
5686.5327	&	1800	&	5807.7326	&	600	&	5658.5870	&	1210	\\
5701.5513	&	1750	&	5844.6708	&	900	&	5660.6107	&	704	\\
5718.4729	&	2000	&	5861.7131	&	900	&	5663.5411	&	1200	\\
5745.4261	&	1950	&	5887.6807	&	900	&	5666.5345	&	1800	\\
5761.5685	&	1800	&	5932.5869	&	900	&	5937.7967	&	1800	\\
5772.5841	&	1950	&	5942.6039	&	1200	&	5964.6900	&	3400	\\
5778.4452	&	1700	&	5949.5190	&	900	&	5965.5843	&	2400	\\
5795.3894	&	1500	&	5954.4967	&	1000	&	5968.6517	&	1800	\\
5808.3721	&	1500	&	5965.4262	&	1100	&	5990.7225	&	1800	\\
5829.3696	&	1500	&	5968.6346	&	900	&	6017.6145	&	1500	\\
5999.7619	&	900	&	5978.5722	&	1050	&	6042.4773	&	1800	\\
6027.6990	&	1500	&	6007.5012	&	820	&	6060.4147	&	1800	\\
6049.5559	&	1500	&	6043.4301	&	900	&	6064.5674	&	1800	\\
6082.5484	&	1150	&	6058.3969	&	1500	&	6066.4652	&	1800	\\
6135.4325	&	800	&	6066.3756	&	1800	&	6068.4574	&	1800	\\
6135.4423	&	800	&	6068.3894	&	1250	&	6082.4504	&	1800	\\
6139.4292	&	1400	&	6069.3668	&	900	&	6138.4374	&	1800	\\
6145.4352	&	1000	&			&		&	6141.4016	&	1800	\\
		&		&			&		&	6196.3415	&	1800	\\
		&		&			&		&	6301.7679	&	1800	\\

\hline
\end{tabular}
\end{table}

\subsection{Radial Velocities}
\citet{Oestensen11} determined preliminary orbital periods of both BD+29$^{\circ}$3070, BD+34$^{\circ}$1543 and Feige\,87 based on the radial velocities of the cool companion and assuming circular orbits, resulting in respectively 1160 $\pm$ 67 days, 818 $\pm$ 21 days and 915 $\pm$ 16 days. These long periods allow us to sum spectra that are taken within a five-day interval to increase the signal to noise (S/N), without significantly smearing or broadening the spectral lines. This five-day interval corresponds to about 0.5 \% of the orbital period, and a maximum radial velocity shift of 0.06 km~s$^{-1}$. After this merging, 28 spectra of BD+29$^{\circ}$3070, 22 spectra of BD+34$^{\circ}$1543 and 19 spectra of Feige\,87 remain, with a S/N varying from 25 to 50. These spectra with the averaged BJD in case of the merged spectra are displayed in Table \ref{tb-rv1}, \ref{tb-rv2} and \ref{tb-rv3}.

The radial velocities of the MS components are determined with the cross-correlation method of the HERMES pipeline, based on a discrete number of line positions. This is possible because the sdB component has only a few H and He lines, which are avoided in the cross correlation. To determine the radial velocities of the MS components of the three systems, a G2-type mask was used on orders 55-74 (4780 - 6530 \AA) as these orders give the best compromise between maximum S/N for G-K type stars and absence of telluric influence. The final errors on the radial velocities (see Table \ref{tb-rv1}, \ref{tb-rv2} and \ref{tb-rv3}) are calculated taking into account the formal errors on the Gaussian fit to the normalized cross-correlation function and the error due to the stability of the wavelength calibration.

To determine the radial velocities of the sdB components a different technique is necessary as these stars have only few spectral lines visible in the composite spectra. The only spectral line that is not contaminated by the MS components is the He\,I blend at 5875.61 \AA. To derive the radial velocity based on only one line, a more specific method is necessary to avoid unacceptably high errors. The region around the He\,I line is first cleaned by hand of all remaining cosmic rays after which it is normalized by fitting low order polynomials to the spectrum. The cleaned and normalized spectra are then cross correlated with a high-resolution synthetic sdB spectrum from the LTE grids of \citet{Heber00}. For all three systems a synthetic spectrum of $T_{\rm{eff}}$ = 30000 K, $\log{g}$ = 5.50 dex, and a resolution matching that of HERMES was used. Spectra with different $T_{\rm{eff}}$ and $\log{g}$ were tried, but did not result in a significant change.
The cross correlation (CC) is carried out in wavelength space, and the resulting radial velocity is calculated by fitting a Gaussian to the cross-correlation function. The error is determined by performing a Monte-Carlo (MC) simulation in which Gaussian noise is added to the observed spectra after which the CC is repeated. The final error is based on the standard deviation of the radial-velocity results of 1000 MC iterations, the wavelength stability of HERMES and the dependence on the used sdB template. For a more elaborate explanation on the derivation of radial velocities from the HERMES spectra see \citetalias[Sect.\,2.1]{Vos12}. The final radial velocities of both the MS and sdB component of BD+29$^{\circ}$3070 together with their errors are given in Table \ref{tb-rv1}, while those of BD+34$^{\circ}$1543 and Feige\,87 can be found in Table \ref{tb-rv2} and \ref{tb-rv3} respectively.

The FOCES spectrum of BD+34$^{\circ}$1543 is analyzed in exactly the same way as the HERMES spectra to determine the radial velocities of both components. The results are given in Table \ref{tb-rv2} together with the HERMES results.

\begin{table}
\caption{The radial velocities of both components of BD+29$^{\circ}$3070.}
\label{tb-rv1}
\centering
\begin{tabular}{lrrrr}
\hline\hline
\noalign{\smallskip}
	&	\multicolumn{2}{r}{MS component}	&	\multicolumn{2}{r}{sdB component}	\\
BJD	&		RV	&	Error		&		RV	&	Error		\\
-2450000	&	km s$^{-1}$	&	km s$^{-1}$	&	km s$^{-1}$	&	km s$^{-1}$	\\\hline
\noalign{\smallskip}
5029.4643	&	-52.88	&	0.26	&	-68.81	&	0.66	\\
5040.4568	&	-52.93	&	0.29	&	-67.11	&	0.54	\\
5057.9327	&	-53.60	&	0.31	&	-67.54	&	0.63	\\
5321.6137	&	-62.78	&	0.24	&	-42.40	&	0.63	\\
5351.4739	&	-63.33	&	0.28	&	-41.13	&	0.70	\\
5425.4024	&	-64.21	&	0.32	&	-38.11	&	0.92	\\
5619.7185	&	-61.96	&	0.26	&	-47.31	&	0.40	\\
5638.6611	&	-61.48	&	0.26	&	-46.01	&	0.60	\\
5648.6732	&	-61.46	&	0.28	&	-49.33	&	0.42	\\
5658.6816	&	-61.23	&	0.28	&	-49.87	&	0.46	\\
5670.5225	&	-60.54	&	0.31	&	-49.68	&	0.65	\\
5678.5495	&	-60.54	&	0.29	&	-49.87	&	0.57	\\
5686.5327	&	-60.53	&	0.28	&	-50.48	&	1.51	\\
5701.5513	&	-60.16	&	0.26	&	-50.08	&	0.45	\\
5718.4729	&	-59.57	&	0.31	&	-51.23	&	0.62	\\
5745.4260	&	-58.64	&	0.27	&	-53.05	&	0.56	\\
5761.5685	&	-57.96	&	0.27	&	-54.34	&	0.47	\\
5772.5840	&	-57.41	&	0.42	&	-55.12	&	1.16	\\
5778.4452	&	-57.97	&	0.27	&	-56.52	&	0.53	\\
5795.3894	&	-57.04	&	0.26	&	-56.18	&	0.52	\\
5808.3721	&	-57.29	&	0.26	&	-57.48	&	0.81	\\
5829.3696	&	-56.31	&	0.27	&	-60.02	&	0.49	\\
5999.7619	&	-53.06	&	0.30	&	-68.54	&	0.60	\\
6027.6990	&	-52.27	&	0.24	&	-69.76	&	0.54	\\
6049.5559	&	-52.56	&	0.28	&	-70.69	&	0.59	\\
6082.5484	&	-51.93	&	0.26	&	-71.97	&	0.52	\\
6136.7680	&	-51.29	&	0.30	&	-71.20	&	0.97	\\
6145.4352	&	-51.17	&	0.27	&	-73.82	&	0.56	\\
\hline
\end{tabular}
\end{table}

\begin{table}
\caption{The radial velocities of both components of BD+34$^{\circ}$1543.}
\label{tb-rv2}
\centering
\begin{tabular}{lrrrr}
\hline\hline
\noalign{\smallskip}
	&	\multicolumn{2}{r}{MS component}	&	\multicolumn{2}{r}{sdB component}	\\
BJD	&		RV	&	Error		&		RV	&	Error		\\
-2450000	&	km s$^{-1}$	&	km s$^{-1}$	&	km s$^{-1}$	&	km s$^{-1}$	\\\hline
\noalign{\smallskip}
1576.5166\tablefootmark{a}	&	29.23	&	0.25	&	38.66   &	1.02	\\
5166.5252	&	37.78	&	0.28	&	22.99	&	1.06	\\
5218.5730	&	38.22	&	0.16	&	23.70	&	0.64	\\
5239.0565	&	37.89	&	0.18	&	22.93	&	0.85	\\
5475.6918	&	28.84	&	0.15	&	40.31	&	0.72	\\
5505.2118	&	27.43	&	0.18	&	39.93	&	0.59	\\
5553.5767	&	26.53	&	0.14	&	42.56	&	0.41	\\
5614.5668	&	26.06	&	0.18	&	43.16	&	0.86	\\
5621.5560	&	26.32	&	0.15	&	43.37	&	0.55	\\
5658.3476	&	26.49	&	0.18	&	42.43	&	1.07	\\
5807.7326	&	29.81	&	0.16	&	37.27	&	0.66	\\
5844.6708	&	30.83	&	0.16	&	36.26	&	0.54	\\
5861.7131	&	31.03	&	0.14	&	35.21	&	0.48	\\
5887.6807	&	32.11	&	0.16	&	33.09	&	0.44	\\
5932.5869	&	33.17	&	0.14	&	30.80	&	0.47	\\
5942.6039	&	33.67	&	0.14	&	31.00	&	0.43	\\
5952.0079	&	33.85	&	0.14	&	31.00	&	0.47	\\
5967.0304	&	34.00	&	0.16	&	30.38	&	0.45	\\
5978.5721	&	34.43	&	0.17	&	28.29	&	0.69	\\
6007.5012	&	35.16	&	0.14	&	26.71	&	0.47	\\
6043.4301	&	36.11	&	0.15	&	25.59	&	0.61	\\
6058.3969	&	36.20	&	0.22	&	25.36	&	1.45	\\
6068.0440	&	36.10	&	0.18	&	24.30	&	0.77	\\
\hline
\end{tabular}
\tablefoot{\tablefoottext{a}{FOCES spectrum}}
\end{table}

\begin{table}
\caption{The radial velocities of both components of Feige\,87.}
\label{tb-rv3}
\centering
\begin{tabular}{lrrrr}
\hline\hline
\noalign{\smallskip}
	&	\multicolumn{2}{r}{MS component}	&	\multicolumn{2}{r}{sdB component}	\\
BJD	&		RV	&	Error		&		RV	&	Error		\\
-2450000	&	km s$^{-1}$	&	km s$^{-1}$	&	km s$^{-1}$	&	km s$^{-1}$	\\\hline
\noalign{\smallskip}
5028.4488	&	28.02	&	0.12	&	42.74	&	0.48	\\
5341.5722	&	31.03	&	0.13	&	39.20	&	0.71	\\
5351.4184	&	31.74	&	0.31	&	37.43	&	0.93	\\
5371.4175	&	32.33	&	0.25	&	35.43	&	0.47	\\
5566.7810	&	40.20	&	0.35	&	22.72	&	0.59	\\
5616.6402	&	40.42	&	0.31	&	22.35	&	0.69	\\
5622.7074	&	39.77	&	0.35	&	20.36	&	0.47	\\
5656.6032	&	39.63	&	0.36	&	21.10	&	0.86	\\
5663.5411	&	39.78	&	0.38	&	20.40	&	0.49	\\
5937.7967	&	29.72	&	0.16	&	41.10	&	0.58	\\
5964.6900	&	28.30	&	0.32	&	42.98	&	0.50	\\
5990.7225	&	26.68	&	0.17	&	46.81	&	0.75	\\
6017.6145	&	24.14	&	0.81	&	46.12	&	0.95	\\
6042.4773	&	24.42	&	0.37	&	49.18	&	0.80	\\
6060.4147	&	24.30	&	0.37	&	51.28	&	1.11	\\
6066.4651	&	24.04	&	0.25	&	50.75	&	0.63	\\
6082.4504	&	23.80	&	0.28	&	50.80	&	0.62	\\
6196.3415	&	26.92	&	0.29	&	45.67	&	0.72	\\
6301.7679	&	32.12	&	0.30	&	36.19	&	0.48	\\
\hline
\end{tabular}
\end{table}

\subsection{Orbital parameters} \label{s-orbitalparam}

\begin{figure*}[!t]
\centering
\includegraphics{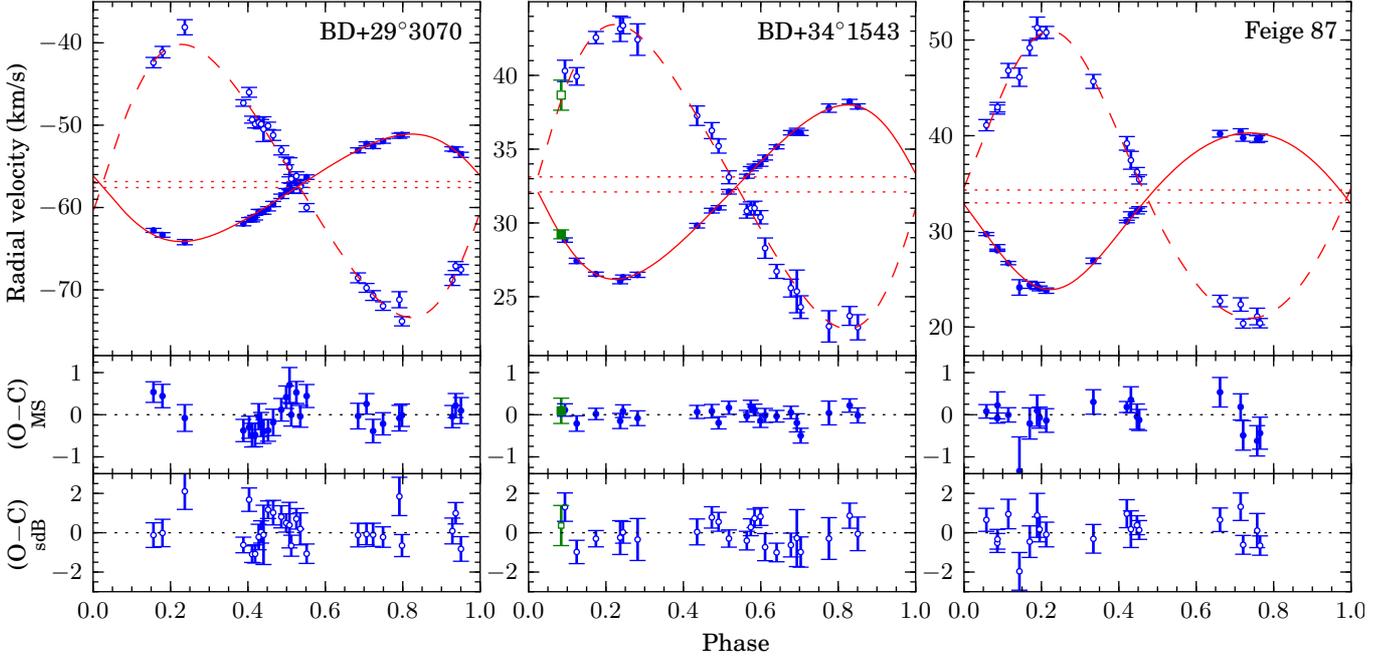}
\caption{The radial velocity curves for BD+29$^{\circ}$3070 (left), BD+34$^{\circ}$1543 (center) and Feige\,87 (right). Top: spectroscopic orbital solution (solid line: MS, dashed line: sdB), and the observed radial velocities (HERMES: blue circles, FOCES: green squares, filled symbols: MS component, open symbols: sdB component). The measured system velocities of both components are shown by a dotted line. Middle: residuals of the MS component. Bottom: residuals of the sdB component.}\label{fig-rvcurves}
\end{figure*}

\begin{table*}
\centering
\caption{Spectroscopic orbital solutions for both the main-sequence (MS) and subdwarf-B (sdB) component of BD+29$^{\circ}$3070 , BD+34$^{\circ}$1543 and Feige\,87.}
\label{tb-specparam}
\begin{tabular}{lrr@{\hskip 1cm}rr@{\hskip 1cm}rr}
\hline\hline
\noalign{\smallskip}
		& \multicolumn{2}{c}{BD+29$^{\circ}$3070} & \multicolumn{2}{c}{BD+34$^{\circ}$1543} & \multicolumn{2}{c}{Feige\,87}\\
Parameter	& \multicolumn{1}{c}{MS} & \multicolumn{1}{c}{sdB} & \multicolumn{1}{c}{MS} & \multicolumn{1}{c}{sdB} & \multicolumn{1}{c}{MS} & \multicolumn{1}{c}{sdB}\\\hline
\noalign{\smallskip}
$P$ (d)           &  \multicolumn{2}{c}{1283 $\pm$ 63}                      &  \multicolumn{2}{c}{972 $\pm$ 2}       &     \multicolumn{2}{c}{936 $\pm$ 2}         \\
$T_0$             &  \multicolumn{2}{c}{2453877  $\pm$ 41}                  &  \multicolumn{2}{c}{2451519  $\pm$ 11} &     \multicolumn{2}{c}{2453259  $\pm$ 21}   \\
$e$               &  \multicolumn{2}{c}{0.15  $\pm$ 0.01}                   &  \multicolumn{2}{c}{0.16  $\pm$ 0.01}  &     \multicolumn{2}{c}{0.11  $\pm$ 0.01}    \\
$\omega$          &  \multicolumn{2}{c}{1.60  $\pm$ 0.22}                   &  \multicolumn{2}{c}{1.58  $\pm$ 0.07}  &     \multicolumn{2}{c}{2.92  $\pm$ 0.15}    \\
$q$               &  \multicolumn{2}{c}{0.39 $\pm$ 0.01}                    &  \multicolumn{2}{c}{0.57 $\pm$ 0.01}   &     \multicolumn{2}{c}{0.55 $\pm$ 0.01}     \\
$\gamma$ (km s$^{-1}$)        &  $-$57.58 $\pm$ 0.36  &  $-$56.8 $\pm$ 0.9  &  32.10 $\pm$ 0.06  &  33.12 $\pm$ 0.15 &     32.98 $\pm$ 0.08  &  34.32 $\pm$ 0.16   \\
$K$  (km s$^{-1}$)            &  6.53 $\pm$ 0.31      &  16.6 $\pm$ 0.6     &  5.91 $\pm$ 0.07   &  10.31 $\pm$ 0.22 &     8.19 $\pm$ 0.11   &  15.01 $\pm$ 0.21   \\
$a$ $\sin{i}$ ($R_{\odot}$)   &  164 $\pm$ 15         &  416 $\pm$ 35       &  112 $\pm$ 2       &  196 $\pm$ 4      &     150 $\pm$ 2       &  276 $\pm$ 3        \\
$M$ $\sin^3{i}$ ($M_{\odot}$) &  1.15 $\pm$ 0.19      &  0.45 $\pm$ 0.08    &  0.27 $\pm$ 0.01   &  0.15 $\pm$ 0.01  &     0.77 $\pm$ 0.03   &  0.42 $\pm$ 0.01    \\
\hline
\end{tabular}
\tablefoot{$a$ denotes the semi-major-axis of the orbit. The quoted errors are the standard deviation from the results of 5000 iterations in a Monte Carlo simulation.}
\end{table*}

The orbital parameters of the sdB and MS components are calculated by fitting a Keplerian orbit to the radial velocity measurements, while adjusting the period ($P$), time of periastron ($T_0$), eccentricity ($e$), angle of periastron ($\omega$), two amplitudes ($K_{\rm{MS}}$ and $K_{\rm{sdB}}$) and two systemic velocities ($\gamma_{\rm{MS}}$ and $\gamma_{\rm{sdB}}$). As a first guess for these parameters, the results of \citet{Oestensen11} were used. The radial velocity measurements were weighted according to their errors as $w = 1/\sigma$. For each system, the \citet{Lucy71} test was used to check if the orbit is significantly eccentric. In the fitting process, the system velocities of both components are allowed to vary independently of each other, to allow for gravitational redshift effects in the sdB component (see \citetalias[Sect. 4]{Vos12}, and Sect. \ref{s-gr} in this paper). The uncertainties on the final parameters are obtained using 5000 iterations in a Monte-Carlo simulation where the radial 
velocities were perturbed based on their errors. 
The spectroscopic parameters of BD+29$^{\circ}$3070 and BD+34$^{\circ}$1543 are shown in Table \ref{tb-specparam}. The radial-velocity curves and the best fits are plotted in Fig. \ref{fig-rvcurves}.

\begin{figure*}[!t]
\centering
\includegraphics{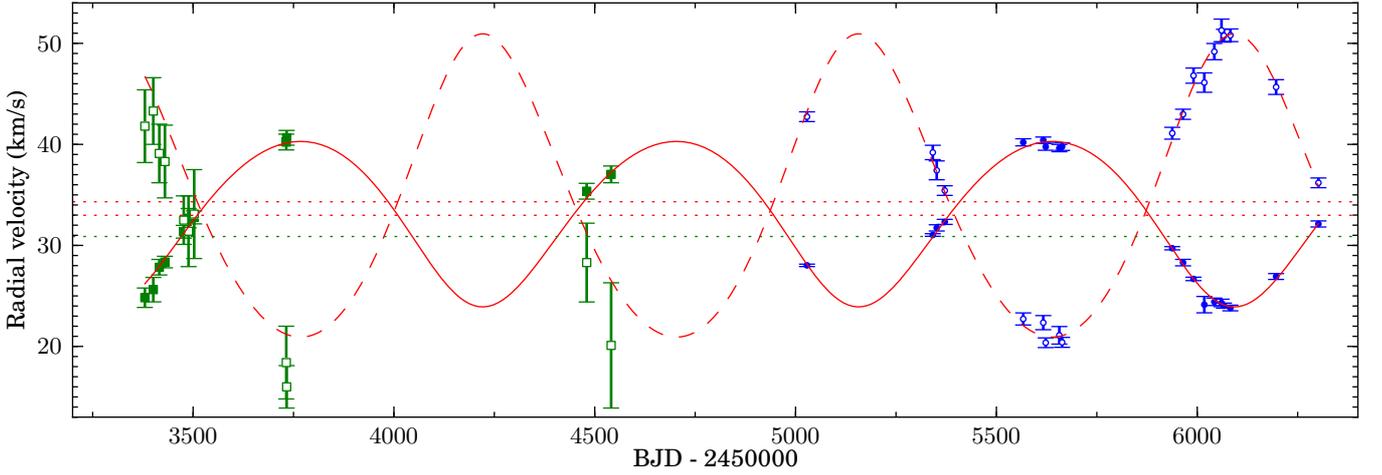}
\caption{The radial velocity curve of Feige\,87, showing the spectroscopic orbital solution (solid line: MS, dashed line: sdB), and the observed radial velocities (HERMES: blue circles, Barlow et al.: green squares, filled symbols: MS component, open symbols: sdB component). The measured system velocities of both components based on the HERMES data and the MS radial velocities of Barlow et al. are shown by a dotted red line, while the system velocity of the sdB component based on the data of Barlow et al. is plotted in a green dotted line. }\label{fig-rvcurvesBarlow}
\end{figure*}

Feige\,87 (= PG\,1338+611) has been studied by \citet{Barlow12} as part of a long term observing program with the Hobby-Eberly telescope lasting from January 2005 till March 2008. They published radial velocities for both the MS and sdB component. However, \citet{Barlow12} find a difference in systemic velocity for the MS and sdB component of $\gamma_{\rm{sdB}} - \gamma_{\rm{MS}} = -2.1 \pm 1.0$ km s$^{-1}$ (compared to $\gamma_{\rm{sdB}} - \gamma_{\rm{MS}} = 1.30 \pm 0.11$ km s$^{-1}$ for the HERMES spectra), which they attribute to gravitational redshift. If this was caused by gravitational redshift, this shift would mean that the MS component has a higher surface gravity than the sdB component, a highly unlikely situation (see also Sect. \ref{s-gr}). A more plausible cause can be found in the lines used to derive the radial velocities of the sdB component. \citet{Barlow12} used both the \ion{He}{i} $\lambda$ 4472 and \ion{He}{i} $\lambda$ 5876 lines, but when comparing the \ion{He}{i} $\lambda$ 4472 line 
with a synthetic 
G2 spectrum, it is clear that this line is significantly contaminated by spectral features of the cool companion.
In the analysis of the radial-velocity curves of Feige\,87, we used their results of \citet{Barlow12} of the MS component, but discarded the results of the sdB component. The phase-folded radial velocity curve of the HERMES data is shown in Fig. \ref{fig-rvcurves}, while the radial velocity curve of both HERMES and \citet{Barlow12} is shown in Fig. \ref{fig-rvcurvesBarlow}. The spectroscopic parameters of Feige\,87 are given in Table \ref{tb-specparam}. 

%
%

\section{Spectral Energy Distribution}\label{s-SED}
The spectral-energy distribution (SED) of the systems can be used to determine the spectral type of the MS and sdB component. We used photometric SEDs which were fitted with model SEDs to determine both the effective temperature and surface gravity of both components.

\subsection{Photometry}
To collect the photometry of both systems the subdwarf database\footnote{http://catserver.ing.iac.es/sddb/} \citep{Oestensen06}, which contains a compilation of data on hot subdwarf stars collected from the literature, is used. These photometric measurements are supplemented with photometry obtained from several other catalogs as listed in Table \ref{tb-photometry}. In total we obtained 17 photometric measurements for BD+29$^{\circ}$3070, 14 for BD+34$^{\circ}$1543 and 11 for Feige\,87. Both accurate photometric measurements at short and long wavelengths are used to establish the contribution of the hot sdB component and the cool MS component.

\begin{table}
\caption{Photometry of BD+29$^{\circ}$3070, BD+34$^{\circ}$1543 and Feige\,87 collected from the literature.}
\label{tb-photometry}
\centering
\begin{tabular}{lrrrr}
\hline\hline
\noalign{\smallskip}
Band    &   Wavelength  &   Width   &  Magnitude    &   Error   \\
    &   \AA     &   \AA &  mag      &   mag \\\hline
\noalign{\smallskip}
\multicolumn{5}{c}{BD+29$^{\circ}$3070}\\
Johnson $U$\tablefootmark{a}		&	3640	&	550	&	10.030	&	0.020	\\
Johnson $B$\tablefootmark{a}    	&	4450	&	940	&	10.600	&	0.020	\\
Johnson $V$\tablefootmark{a}    	&	5500	&	880	&	10.420	&	0.020	\\
Johnson $B$\tablefootmark{b}    	&	4450	&	940	&	10.590	&	0.013	\\
Johnson $V$\tablefootmark{b}    	&	5500	&	880	&	10.416	&	0.013	\\
Johnson $V$\tablefootmark{c}    	&	5500	&	880	&	10.376	&	0.052	\\
Cousins $I$\tablefootmark{c}    	&	7880	&	1490	&	10.026	&	0.053	\\
Geneva $U$\tablefootmark{a}   		&	3460	&	170	&	10.264	&	0.010	\\
Geneva $B$\tablefootmark{a} 		&	4250	&	283	&	9.638	&	0.010	\\
Geneva $V$\tablefootmark{a} 		&	5500	&	298	&	10.359	&	0.010	\\
Geneva $B1$\tablefootmark{a} 		&	4020	&	171	&	10.523	&	0.011	\\
Geneva $B2$\tablefootmark{a} 		&	4480	&	164	&	11.103	&	0.011	\\
Geneva $V1$\tablefootmark{a} 		&	5400	&	202	&	11.075	&	0.011	\\
Geneva $G$\tablefootmark{a} 		&	5810	&	206	&	11.476	&	0.011	\\
2MASS $J$\tablefootmark{d}  		&	12410	&	1500	&	9.773	&	0.020	\\
2MASS $H$\tablefootmark{d}  		&	16500	&	2400	&	9.621	&	0.023	\\
2MASS $K_s$\tablefootmark{d}  		&	21910	&	2500	&	9.546	&	0.022	\\
\noalign{\smallskip}
\multicolumn{5}{c}{BD+34$^{\circ}$1543}\\
Johnson $B$\tablefootmark{b}	&	4450	&	940	&	10.293	&	0.014	\\
Johnson $V$\tablefootmark{b}	&	5500	&	880	&	10.145	&	0.013	\\
Johnson $V$\tablefootmark{c}	&	5500	&	880	&	10.156	&	0.048	\\
Cousins $I$\tablefootmark{c}	&	7880	&	1490	&	9.758	&	0.077	\\
Geneva $U$\tablefootmark{a}   	&	3460	&	170	&	9.759	&	0.008	\\
Geneva $B$\tablefootmark{a} 	&	4250	&	283	&	9.362	&	0.009	\\
Geneva $V$\tablefootmark{a} 	&	5500	&	298	&	10.140	&	0.007	\\
Geneva $B1$\tablefootmark{a} 	&	4020	&	171	&	10.215	&	0.010	\\
Geneva $B2$\tablefootmark{a} 	&	4480	&	164	&	10.865	&	0.010	\\
Geneva $V1$\tablefootmark{a} 	&	5400	&	202	&	10.854	&	0.010	\\
Geneva $G$\tablefootmark{a} 	&	5810	&	206	&	11.260	&	0.010	\\
2MASS $J$\tablefootmark{d}  	&	12410	&	1500	&	9.485	&	0.023	\\
2MASS $H$\tablefootmark{d}  	&	16500	&	2400	&	9.326	&	0.033	\\
2MASS $K_s$\tablefootmark{d} 	&	21910	&	2500	&	9.207	&	0.018	\\
\noalign{\smallskip}
\multicolumn{5}{c}{Feige\,87}\\
Johnson $B$\tablefootmark{e}	&	4450	&	940	&	11.598	&	0.050	\\
Johnson $V$\tablefootmark{e}	&	5500	&	880	&	11.693	&	0.050	\\
Cousins $R$\tablefootmark{e}	&	6470	&	1515	&	11.671	&	0.050	\\
Cousins $I$\tablefootmark{e}	&	7880	&	1490	&	11.623	&	0.050	\\
Stromgren $U$\tablefootmark{f}	&	3500	&	300	&	11.800	&	0.045	\\
Stromgren $V$\tablefootmark{f}	&	4110	&	190	&	11.747	&	0.045	\\
Stromgren $B$\tablefootmark{f}	&	4670	&	180	&	11.698	&	0.045	\\
Stromgren $Y$\tablefootmark{f}	&	5470	&	230	&	11.730	&	0.090	\\
2MASS $J$\tablefootmark{d}  	&	12410	&	1500	&	11.484	&	0.022	\\
2MASS $H$\tablefootmark{d}  	&	16500	&	2400	&	11.359	&	0.026	\\
2MASS $K_s$\tablefootmark{d}	&	21910	&	2500	&	11.312	&	0.022	\\
\hline
\end{tabular}
\tablebib{
\tablefoottext{a}{\citet{Mermilliod97}}
\tablefoottext{b}{\citet{Kharchenko01}}
\tablefoottext{c}{\citet{Richmond07}}
\tablefoottext{d}{\citet{Cutri2003}}
\tablefoottext{e}{\citet{Allard94}}
\tablefoottext{f}{\citet{Bergeron84}}
}
\end{table}

\subsection{SED fitting}
The SED fitting method is similar to the one described in \citetalias{Vos12}. The observed photometry is fitted with a synthetic SED integrated from model atmospheres. For the MS component Kurucz atmosphere models \citep{Kurucz79} ranging in effective temperature from 4000 to 9000 K, and in surface gravity from $\log{g}$=3.0 dex (cgs) to 5.0 dex (cgs) are used. For the hot sdB component TMAP (T\"{u}bingen NLTE Model-Atmosphere Package, \citealt{Werner03}) atmosphere models with a temperature range from 20000 K to 50000 K, and $\log{g}$ from 4.5 dex (cgs) to 6.5 dex (cgs) are used.

The SEDs are fitted in two steps. First the grid based approach described in \citet{Degroote11} extended for binarity is used to scan the entire parameter space. For each point the $\chi^2$ is calculated as
\begin{equation}
 \chi^2 = \sum_i{ \frac{(O_i-C_i)^2}{\xi_i^2}},
\end{equation}
where $O_i$ is the observed photometry and $C_i$ is the calculated model photometry. The grid point with the lowest $\chi^2$ is used as starting point for a least-squares minimizer which will then determine the final result. In a binary system, there are eight parameters to consider: the effective temperatures ($T_{\rm{eff,MS}}$ and $T_{\rm{eff,sdB}}$), surface gravities ($g_{\rm{MS}}$ and $g_{\rm{sdB}}$) and radii ($R_{\rm{MS}}$ and $R_{\rm{sdB}}$) of both components, the interstellar reddening $E(B-V)$ and the distance ($d$) to the system. The interstellar reddening is presumed equal for both the MS and the sdB component. To increase the accuracy, the models are first corrected for interstellar reddening and then integrated over the photometric pass-bands using the reddening law of \citet{Fitzpatrick2004} with $R_V$ = 3.1. The distance to the system is used as a scale factor and is calculated analytically, by shifting the synthetic models to the photometric observations.

As shown in \citetalias{Vos12}, the mass ratio obtained from the radial velocity curves can be used to couple the radii of both components to their surface gravity, and thus reducing the number of free parameters from eight to six. The total flux of a binary system is then calculated using:
\begin{equation}
 F_{tot}(\lambda) = \frac{G\ M_{\rm{MS}}}{d^2\ g_{\rm{MS}}} \left( F_{\rm{MS}}(\lambda) + q \frac{g_{\rm{MS}}}{g_{\rm{sdB}}}\ F_{\rm{sdB}}(\lambda) \right).
\end{equation}

The uncertainties on the final parameters are determined by calculating two-dimensional confidence intervals (CI) for all parameter pairs. This is done by creating a 2D-grid for each parameter pair. For each point in this grid, the two parameters for which the CI is calculated are kept fixed on the grid-point value, while the least-squares minimizer is used to find the best-fitting values for all other parameters. The resulting $\chi^2$ for each point in this grid is stored. All these $\chi^2$s are then rescaled so that the $\chi^2$ of the best fit has the expected value $k = \rm{N}_{\rm{obs}} - \rm{N}_{\rm{free}}$, with N$_{\rm{obs}}$ the number of observations and N$_{\rm{free}}$ the number of free parameters in the fit. The cumulative density function (CDF) is used to calculate the probability of a model to obtain a certain $\chi^2$ value as:
\begin{equation}
 F(\chi^2,k) = P\left(\frac{k}{2}, \frac{\chi^2}{2}\right) \label{e-cdf}
\end{equation}
Where P is the regularized $\Gamma$-function. Based on the obtained probability distribution, the uncertainties on the parameters can be derived.

\subsection{Results}\label{s-SED-results}

\begin{figure*}[!t]
\centering
\includegraphics{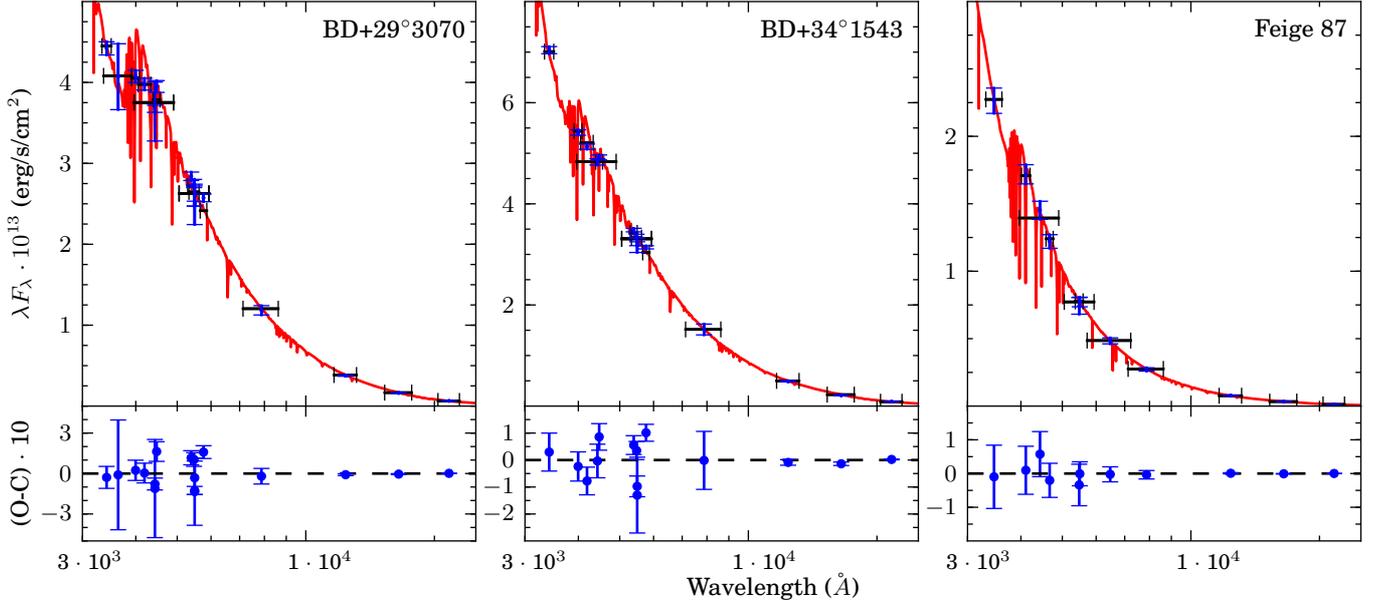}
\caption{The spectral energy distribution and (O-C) of BD+29$^{\circ}$3070 (left), BD+34$^{\circ}$1543 (center) and Feige\,87 (right). The measurements are given in blue, the integrated synthetic models are shown in black, where a horizontal error bar indicates the width of the pass-band. The best fitting model is plotted in red. In the bottom panels the residuals are plotted for each system.}\label{fig-sed}
\end{figure*}

\begin{table*}
\caption{The results of the SED fit for BD+29$^{\circ}$3070, BD+34$^{\circ}$1543 and Feige\,87, together with the 95\% probability intervals derived from the confidence intervals plotted in Fig. \ref{fig-ci}.}
\label{tb-sedresults}
\centering
\begin{tabular}{lrr@{ -- }lrr@{ -- }lrr@{ -- }l}
\hline\hline
\noalign{\smallskip}
		& \multicolumn{3}{c}{BD+29$^{\circ}$3070} & \multicolumn{3}{c}{BD+34$^{\circ}$1543} & \multicolumn{3}{c}{Feige\,87} \\
Parameter	& Best fit & \multicolumn{2}{c}{95\%} & Best fit & \multicolumn{2}{c}{95\%} & Best fit & \multicolumn{2}{c}{95\%} \\\hline
\noalign{\smallskip}
T$_{\rm{eff, MS}}$ (K)		&	6570	&	5800	&	7100	&	6210	&	6000	&	6440	&	5840	&	5300	&	6400	\\
$\log{g}_{\rm{MS}}$ (dex)	&	4.40	&	4.00	&	4.80	&	4.19	&	4.05	&	4.35	&	4.40	&	4.15	&	4.60	\\
T$_{\rm{eff, sdB}}$ (K)		&	28500	&	24000	&	36000	&	36700	&	30000	&	/	&	27400	&	21000	&	33000	\\
$\log{g}_{\rm{sdB}}$ (dex)	&	5.76	&	5.20	&	6.20	&	5.92	&	5.75	&	6.05	&	5.54	&	5.20	&	5.80	\\
E($B-V$)			&	0.007	&	0	&	0.052	&	0.007	&	0	&	0.068	&	0.012	&	0	&	0.057	\\
\hline
\end{tabular}
\tablefoot{For some parameters the confidence intervals could not be calculated due to the limited range of the model atmospheres.}
\end{table*}

\begin{figure*}[!t]
\centering
\includegraphics{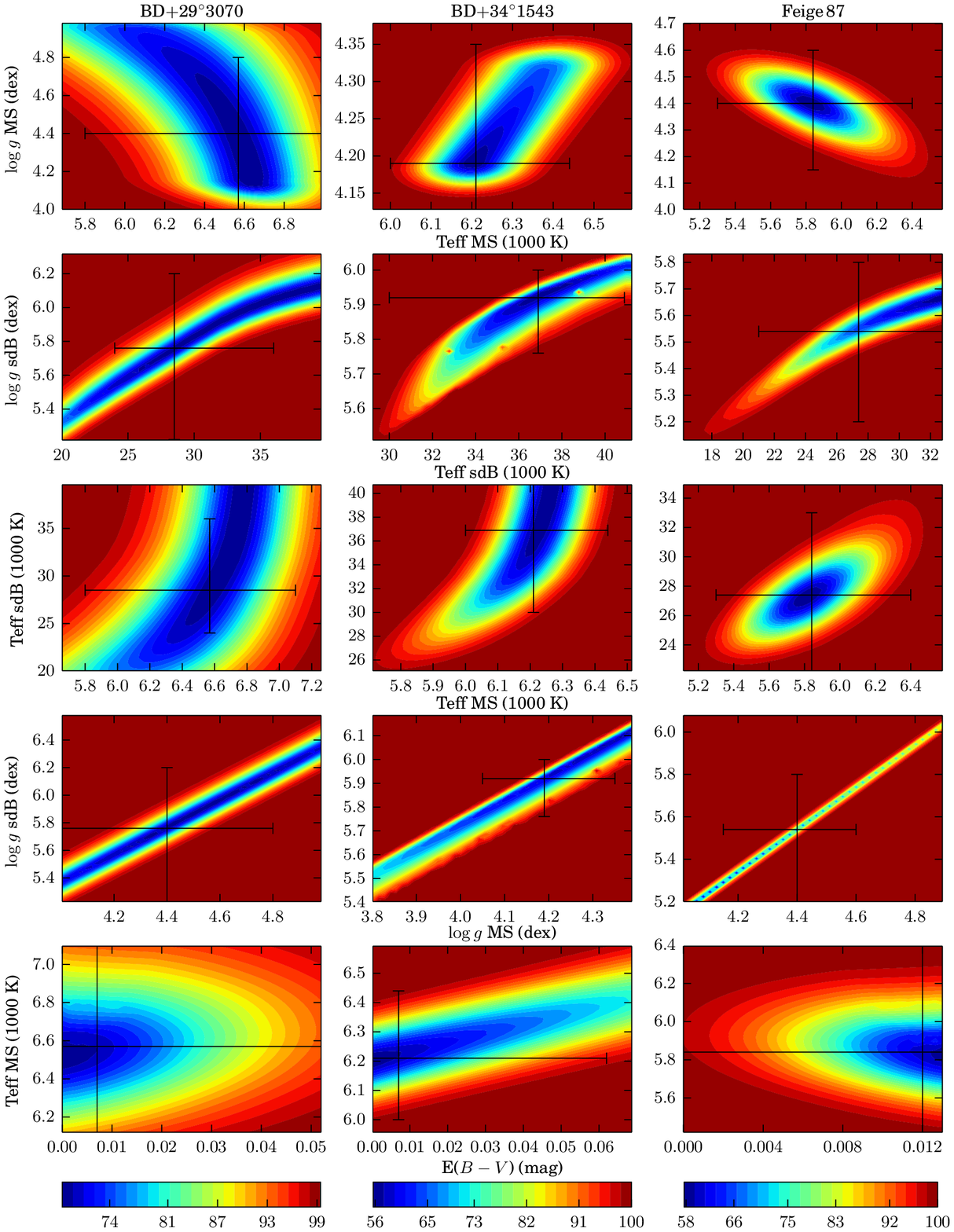}
\caption{The confidence intervals of the SED fits of BD+29$^{\circ}$3070, BD+34$^{\circ}$1543 and Feige\,87. The different colors show the cumulative density probability connected to the $\chi^2$ statistics given by equation \ref{e-cdf}. The best fitting values and there errors are plotted in black.}\label{fig-ci}
\end{figure*}

To fit the SEDs of the three systems, first a grid of composite binary spectra is calculated for ~1\,000\,000 points randomly distributed over the $T_{\rm{eff}}$, $\log{g}$ and E($B-V$) intervals for both components. Then the 50 best-fitting grid points are used as starting values for the least-squares minimizer. For this best fit the CIs are estimated, and used to limit the original ranges on the parameters, after which the fitting process (grid search and least-squares minimizer) is repeated. To determine the CIs of the parameters, for each two parameters a grid with a resolution of 45 $\times$ 45 points was used, with the limits adjusted based on the resulting confidence intervals. When the model atmospheres allowed, the limits of the grid are set to include the 95\,\% CI, when this was not possible, the limits of the model atmosphere grid were used.
The final uncertainties on the parameters are an average of the 95\,\% CIs for that parameter in all two-dimensional CIs that contain that parameter.

The photometry of BD+29$^{\circ}$3070 has a rather large spread, which will result in large uncertainties on the derived parameters. The best fitting effective temperatures are $T_{\rm{eff, MS}}$ = 6570 $\pm$ 550 K and $T_{\rm{eff, sdB}}$ = 28500 $\pm$ 5000 K, with surface gravities of $\log{g}_{\rm{MS}}$ = 4.4 $\pm$ 0.5 dex and $\log{g}_{\rm{sdB}}$ = 5.76 $\pm$ 0.5 dex. The reddening is E($B-V$) = 0.009$_{-0.009}^{+0.041}$ mag. This value is consistent with the maximum reddening E($B-V$)$_{\rm{max}}$ = 0.052 derived from the dust maps of \citet{Schlegel98}. The $\chi^2$ of the best fit is 45.6 which is more than three times as high as the expected value (N$_{\rm{obs}}$ - N$_{\rm{free}}$ = 12), indicating that the stated errors on the photometry are too small. When calculating the confidence intervals, a scaling factor of 3.8 is used when converting the $\chi^2$ values to probabilities. 
As can be seen on the plots of the CIs in Fig. \ref{fig-ci}, there is no strong constraint possible on the surface gravity of the components. Increasing the surface gravity of one component can be countered by increasing the $\log{g}$ of the other component as well, and thus effectively shrinking the radii of both components. For the sdB component there is a strong correlation between $T_{\rm{eff}}$ and $\log{g}$ visible. The effect on the atmosphere models of an increase in $T_{\rm{eff}}$ can be diminished by increasing $\log{g}$, and thus decreasing the radius. However, it is possible to provide an upper limit on the effective temperature as the \ion{He}{ii} lines are not visible in the HERMES spectra, indicating that the effective temperature is below 35000 K \citep{Heber09}.

For BD+34$^{\circ}$1543 fourteen photometric measurements are available, with a smaller spread than for BD+29$^{\circ}$3070. The effective temperatures of the components derived from the SED fit are $T_{\rm{eff, MS}}$ = 6210 $\pm$ 250 K and $T_{\rm{eff, sdB}}$ = 36700 $\pm$ 5000 K, while a surface gravity of $\log{g}_{\rm{MS}}$ = 4.19 $\pm$ 0.20 dex and $\log{g}_{\rm{sdB}}$ = 5.92 $\pm$ 0.40 dex are found. The reddening is determined to be E($B-V$) = 0.007$_{-0.007}^{+0.061}$ which is consistent with the maximum reddening found on the dust maps of \citet{Schlegel98}, and supported by the absence of sharp interstellar absorption lines in the spectrum.
As can be seen in Fig. \ref{fig-ci} the probability distributions of the MS components parameters form a Gaussian-like pattern, and the $T_{\rm{eff}}$ and $\log{g}$ of the MS component have stronger constraints than for BD+29$^{\circ}$3070, independently of the parameters for the sdB components. The uncertainty on the effective temperature and surface gravity of the sdB component is larger as both parameters are correlated in the same way as for BD+29$^{\circ}$3070. However, the presence of clearly visible \ion{He}{i} and \ion{He}{ii} lines in the HERMES spectra of BD+34$^{\circ}$1543 indicates that the effective temperature should be between 35000 and 50000 K.

In the case of BD+34$^{\circ}$1543 the parallax was measured by Hipparcos \citep{Vanleeuwen07} to be 4.22 $\pm$ 1.72 mas. This parallax could be used to fix the distance to the system in the SED fitting process. Since the parallax is of the same order as the projected size of the orbit for this system, one may presume that the Hipparcos parallax is unreliable, but as both components of BD+34$^{\circ}$1543 have a very similar flux in the Hipparcos pass band, the center of light does not change during the orbit. However, the uncertainty on the parallax is large ($\sim$ 40\,\%). Using this parallax to fix the distance to the system, without propagating the uncertainty on it, results in effective temperatures of $T_{\rm{eff, MS}}$ = 6230 $\pm$ 100 K and $T_{\rm{eff, sdB}}$ = 37900 $\pm$ 3500 K, and surface gravities of $\log{g}_{\rm{MS}}$ = 4.04 $\pm$ 0.05 dex and $\log{g}_{\rm{sdB}}$ = 5.81 $\pm$ 0.15 dex. The reddening is determined to be E($B-V$) = 0.011$_{-0.011}^{+0.057}$. The uncertainties are much 
smaller as when the distance is included as a free parameter, especially on the surface gravity. However, due to the large uncertainty on the parallax, only the atmospheric parameters derived with the distance as a free parameter are used.


For Feige\,87 there are only eleven photometric measurements found in the literature, but there is a very good agreement between all the measurements in the different bands. When checking the residuals of the fit, Feige\,87 has the smallest spread of all three systems, with a total $\chi^2$ of 2.9. The SED results in an effective temperature of $T_{\rm{eff, MS}}$ = 5840 $\pm$ 500 K and $T_{\rm{eff, sdB}}$ = 27400 $\pm$ 5000 K for the MS and sdB component, together with a surface gravity of $\log{g}_{\rm{MS}}$ = 4.40 $\pm$ 0.30 dex and $\log{g}_{\rm{sdB}}$ = 5.50 $\pm$ 0.50 dex. The reddening of the system is found to be E($B-V$) = 0.012$_{-0.012}^{+0.045}$, consistent with the results from the dust maps of \citet{Schlegel98}. On Fig. \ref{fig-ci} the probability distributions show a similar pattern as for BD+29$^{\circ}$3070. Although the distribution in $T_{\rm{eff, MS}}$ -- $\log{g}_{\rm{MS}}$ and $T_{\rm{eff, sdB}}$ -- $T_{\rm{eff, MS}}$ have a clearer Gaussian 
pattern, indicating that they are determined more accurate as for BD+29$^{\circ}$3070.

The advantage of having an estimate of the distance to the target is clear when considering the uncertainties of the atmospheric parameters. Because the radii of the components are derived from their surface gravity, limiting the distance will also limit the radii. If the distance is accurately known, the accuracy of the surface gravity can be increased by a factor ten compared to when the distance is treated as a free parameter. Determining atmospheric parameters from photometry will greatly benefit from the Gaia mission that will derive accurate distances of about a billion stars.

\section{Spectral analysis}\label{s-specanal}
The atmospheric parameters determined from the SEDs can be checked using the spectra. The HERMES echelle spectra are not easy to flux calibrate accurately, and therefore not well suited to fit model atmospheres, but it is possible to subtract the continuum contribution of the sdB component as explained in the following subsection. The resulting spectra of the MS components can be used to derive atmospheric parameters based on the \ion{Fe}{i} and \ion{Fe}{ii} lines. Apart from the HERMES spectra, we have obtained flux calibrated long-slit spectra with the Bok telescope (see Sect. \ref{s-specfit}). The resolution of these spectra is too low to determine radial velocities, but they can be used to fit model atmospheres and provide an independent set of atmospheric parameters.

\subsection{Disentangling}
When the spectroscopic parameters of both components in a system are known, it is possible to extract the spectrum of the MS component from the combined spectrum. This is done by subtracting a synthetic sdB spectrum with a surface gravity and effective temperature determined from the SED fit. As the sdB component only has a few lines and only wavelength regions that don't contain balmer or He lines are used, this is equivalent to subtracting the continuum contribution of the sdB component. 

\begin{figure*}[!t]
\centering
\includegraphics{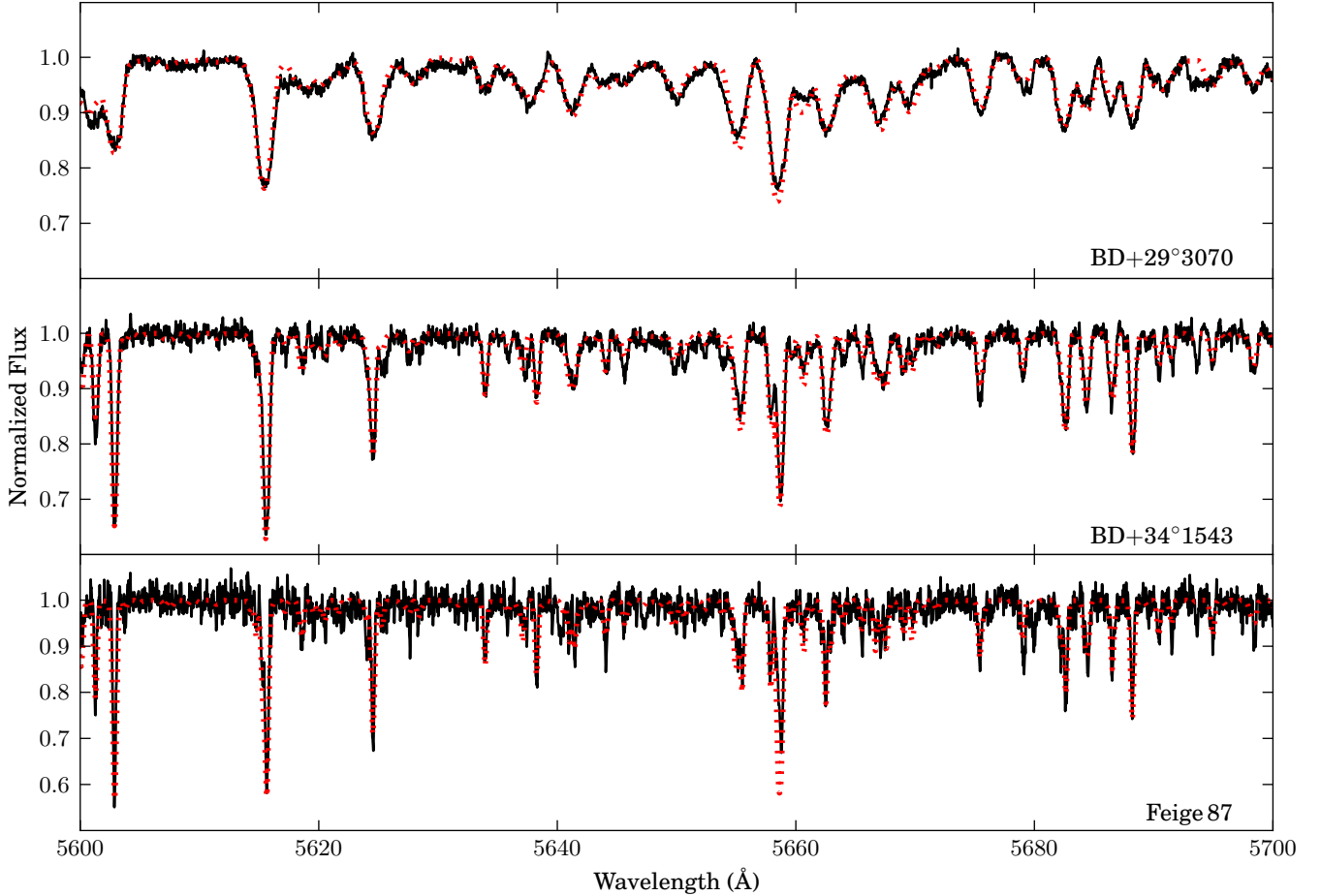}
\caption{The disentangled HERMES spectra of the MS components obtained by subtracting the sdB continuum and averaging all spectra (black full line), together with Kurucz model atmospheres based on the atmospheric parameters determined from the iron lines (red dotted line). Top: BD+29$^{\circ}$3070, center: BD+34$^{\circ}$1543, bottom: Feige\,87.}\label{fig-spectrum}
\end{figure*}

Each HERMES spectrum is treated separately. First the HERMES response curve is removed from the spectrum after which its continuum is determined by fitting a low degree polynomial to the spectrum. Then the continuum contribution of the sdB component is subtracted following:
\begin{equation}
 F_{\rm{tot}}(\lambda) = F_{\rm{MS}}(\lambda) \frac{C(\lambda)}{1+F_{\rm{rat}}(\lambda)} + F_{\rm{sdB}}(\lambda) \frac{C(\lambda) \cdot F_{\rm{rat}}(\lambda)}{1+F_{\rm{rat}}(\lambda)}
\end{equation}
Where $F_{\rm{tot}}$ is the total flux in the HERMES spectrum, $F_{\rm{MS}}$ and $F_{\rm{sdB}}$ are the normalized fluxes of the MS and sdB components, $C$ is the continuum of the HERMES spectrum, and $F_{\rm{rat}}$ is the ratio of the sdB flux to the MS flux. As only the regions where the sdB component does not have significant lines are used, $F_{\rm{sdB}}(\lambda)$ = 1 for every $\lambda$. The normalized MS spectrum is then given by:
\begin{equation}
 F_{\rm{MS}}(\lambda) =  \left( F_{\rm{tot}}(\lambda) - \frac{C(\lambda) \cdot F_{\rm{rat}}(\lambda)}{1+F_{\rm{rat}}(\lambda)} \right) / \left( \frac{C(\lambda)}{1+F_{\rm{rat}}(\lambda)} \right).
\end{equation}
The obtained MS spectra are shifted to zero velocity, and averaged weighted by their S/N. An example region of the final extracted spectra for both systems is plotted in Fig. \ref{fig-spectrum}.

This way of disentangling the spectra is dependent on the atmospheric parameters of both components, which means that the disentangling process needs to be repeated when a new set of atmospheric parameters for the MS component is derived from the iron lines. This is done until convergence, which happened for all systems after two iterations. The parameters of the sdB component cannot be derived from the resulting spectra, but tests were performed to check their influence. As it turns out, changing the $T_{\rm{eff}}$ or $\log{g}$ of the sdB component within the errors determined in the SED fitting process, does not have a significant effect on the parameters of the MS component derived from the iron lines.

\subsection{Atmospheric parameters and abundances}
The Versatile Wavelength Analysis\footnote{https://sites.google.com/site/vikingpowersoftware/} (VWA) tool developed by \citet{Bruntt02} is used to determine the effective temperature, surface gravity, microturbulent velocity and abundances of the MS components. VWA generates synthetic spectra using the SYNTH software of \citet{Valenti96}. Atomic line data is taken from the VALD database \citep{Kupka99}, but the $\log(gf)$ values are adjusted so that every line measured by \citet{Wallace98} reproduces the atmospheric abundances by \citet{Grevesse07}. The atmosphere models are interpolated from MARCS model atmospheres \citep{Gustafsson08} using the solar composition of \citet{Grevesse07}. The VWA package fits abundances in a semi-automatic way. It first selects the least blended lines in the spectra, and determines the abundances of these lines by calculating synthetic spectra for each line while iteratively changing the input abundance until the equivalent widths of the observed and synthetic spectrum 
match. The main advantage of VWA is that the synthetic spectrum includes the contribution of neighboring lines, thus making it possible to analyze stars with a high rotational velocity. A detailed description of the VWA software can be found in \citet{Bruntt04, Bruntt08, Bruntt09, Bruntt10a, Bruntt10b}.

Before the spectra are analyzed, they are carefully normalized with the {\tt RAINBOW} tool of the VWA package. Then the spectra are compared to synthetic spectra, and projected rotational velocities of 52 $\pm$ 5 km s$^{-1}$, 17 $\pm$ 4 km s$^{-1}$ and 8 $\pm$ 3 km s$^{-1}$ are found for the G-star components of respectively BD+29$^{\circ}$3070, BD+34$^{\circ}$1543 and Feige\,87. Especially BD+29$^{\circ}$3070 has a high rotational velocity, resulting in severe line blending, which makes it difficult to derive the atmospheric parameters. These parameters are determined using only the iron lines. The effective temperature is determined by requiring the abundance of the \ion{Fe}{i} lines to be independent of the excitation potential. The surface gravity is derived by requiring the same abundance for \ion{Fe}{i} and \ion{Fe}{ii} lines, and checked by fitting synthetic spectra to several calcium and magnesium lines that are sensitive to changes in $\log{g}$ \citep{Gray2005}. Furthermore the independence 
of 
abundance on equivalent 
width gives the microturbulence velocity. When the atmospheric parameters are determined the final abundances of all measured lines are calculated, and the overall metallicity is obtained by averaging all abundances over the measured elements weighted by the number of lines found for each element.

The disentangled spectrum of BD+29$^{\circ}$3070 has the highest signal-to-noise ratio of the three systems (S/N $\sim$ 130), making it possible to derive robust parameters regardless of the high rotational blending. In total, 401 suitable lines were selected after comparing the data with a synthetic spectrum. In the abundance determination process 56 lines (of which 34 \ion{Fe}{i} and 3 \ion{Fe}{ii} lines) were used and had an equivalent width between 10 and 90 m\AA. The three \ion{Fe}{ii} lines have too large scatter to constrain the surface gravity and are ignored for this purpose. To obtain an estimate of the surface gravity three calcium lines (Ca $\lambda$ 6122, Ca $\lambda$ 6162, Ca $\lambda$ 6439) and the magnesium triplet (Mg-1b $\lambda$ 5172) are fitted with synthetic spectra. This resulted in a surface gravity of $\log{g}$ = 4.3 $\pm$ 0.5 dex. Using the surface gravity determined from the Ca and Mg lines, the atmospheric parameters determined based on the \ion{Fe}{i} lines are: $T_{\rm{eff}}$ = 
6100 $\pm$ 200 K and $v_{\rm{micro}}$ = 1.50 $\pm$ 0.35 km s$^{-1}$. The overall metallicity calculated using all 56 lines is [M/H] = 0.09 $\pm$ 0.21. The abundances of all lines are given in Table \ref{tb-abundances}.

Having a lower rotational velocity and a still a high signal to noise (S/N $\sim$ 90), more useful lines (578) were found in BD+34$^{\circ}$1543. After the fitting process 99 lines (including 50 \ion{Fe}{i} and 6 \ion{Fe}{ii} lines) were fitted and had an equivalent width between 10 and 90 m\AA. In this case the scatter on the abundances of the \ion{Fe}{ii} lines was low, we used it together with the \ion{Fe}{i} lines to constrain the surface gravity, resulting in $\log{g}$ = 4.2 $\pm$ 0.3 dex. This value is supported by fitting synthetic spectra to the same Ca and Mg lines as for BD+29$^{\circ}$3070. Based on the \ion{Fe}{i} lines an effective temperature of $T_{\rm{eff}}$ = 6150 $\pm$ 150 K and a microturbulence of $v_{\rm{micro}}$ = 1.45 $\pm$ 0.25 km s$^{-1}$ were found. The averaged metallicity is found to be slightly sub solar at [M/H] = $-$0.24 $\pm$ 0.12. The abundances of all elements are shown in Table \ref{tb-abundances}. 

Feige\,87 has the lowest rotational velocity, but this advantage is partly countered by the low signal-to-noise ratio of the disentangled spectrum (S/N $\sim$ 35). From the 428 selected lines 93 were used (including 51 \ion{Fe}{i} lines and 3 \ion{Fe}{ii} lines). The scatter of the \ion{Fe}{ii} lines is high compared to the previous system, thus they are not very reliable in constraining the surface gravity. Based on fitting synthetic spectra to Ca and Mg lines, and on the ionization balance between \ion{Fe}{i} and \ion{Fe}{ii}, we find a surface gravity of $\log{g}$ = 4.5 $\pm$ 0.6 dex. Based on the \ion{Fe}{i} lines, the effective temperature is constrained to $T_{\rm{eff}}$ = 6175 $\pm$ 150 K and the microturbulence results in $v_{\rm{micro}}$ = 1.15 $\pm$ 0.25 km s$^{-1}$. The average metallicity based on 96 lines in total is clearly sub solar at [M/H] = $-$0.48 $\pm$ 0.26. The abundances of all elements are shown in Table \ref{tb-abundances}. 

\begin{table}
\centering
\caption{The abundances ([El/H]) for the MS components of BD+29$^{\circ}$3070, BD+34$^{\circ}$1543 and Feige\,87 obtained from the disentangled HERMES spectra.}
\label{tb-abundances}
\begin{tabular}{lr@{ $\pm$ }l@{\hspace{0.2cm}}lr@{ $\pm$ }l@{\hspace{0.2cm}}lr@{ $\pm$ }l@{\hspace{0.2cm}}l}
\hline\hline
\noalign{\smallskip}
           & \multicolumn{3}{c}{BD+29$^{\circ}$3070} & \multicolumn{3}{c}{BD+34$^{\circ}$1543} &   \multicolumn{3}{c}{Feige\,87}\\
Ion & \multicolumn{2}{c}{[El/H]} & N\tablefootmark{a} & \multicolumn{2}{c}{[El/H]} & N\tablefootmark{a} & \multicolumn{2}{c}{[El/H]} & N\tablefootmark{a} \\
\hline
{Al \sc   i}	&	0.26	&	0.1	&	2	&  \multicolumn{2}{c}{ }	&		&  \multicolumn{2}{c}{ }	&		\\
{Si \sc   i}	&	0.24	&	0.19	&	8	&	$-$0.12	&	0.08	&	11	&	$-$0.38	&	0.13	&	9	\\
{Si \sc  ii}	&  \multicolumn{2}{c}{0.25}	&	1	&	$-$0.21	&	0.06	&	2	&  \multicolumn{2}{c}{ }	&		\\
{Ca \sc   i}	&  \multicolumn{2}{c}{0.03}	&	1	&	$-$0.14	&	0.04	&	7	&	$-$0.27	&	0.06	&	7	\\
{Sc \sc  ii}	&  \multicolumn{2}{c}{ }	&		&  \multicolumn{2}{c}{$-$0.09}	&	1	&	$-$0.70	&	0.13	&	3	\\
{Ti \sc   i}	&  \multicolumn{2}{c}{ }	&		&	$-$0.05	&	0.06	&	3	&	0.09	&	0.25	&	2	\\
{Ti \sc  ii}	&  \multicolumn{2}{c}{ }	&		&	$-$0.07	&	0.12	&	2	&  \multicolumn{2}{c}{$-$0.77}	&	1	\\
{Cr \sc   i}	&  \multicolumn{2}{c}{ }	&		&	$-$0.27	&	0.13	&	4	&	$-0.51$	&	0.04	&	2	\\
{Cr \sc  ii}	&  \multicolumn{2}{c}{0.08}	&	1	&  \multicolumn{2}{c}{$-$0.25}	&	1	&  \multicolumn{2}{c}{ }	&		\\
{Mn \sc   i}	&  \multicolumn{2}{c}{ }	&		&  \multicolumn{2}{c}{$-$0.07}	&	1	&  \multicolumn{2}{c}{$-$0.82}	&	1	\\
{Mg \sc   i}	&  \multicolumn{2}{c}{ }	&		&  \multicolumn{2}{c}{ }	&		&  \multicolumn{2}{c}{$-$0.12}	&	1	\\
{Fe \sc   i}	&	0.04	&	0.16	&	34	&	$-$0.26	&	0.07	&	50	&	$-$0.49	&	0.17	&	51	\\
{Fe \sc  ii}	&	0.1	&	0.18	&	3	&	$-$0.26	&	0.12	&	6	&	$-$0.60	&	0.4	&	3	\\
{Ni \sc   i}	&	0.08	&	0.16	&	6	&	$-$0.34	&	0.10	&	11	&	$-$0.40	&	0.25	&	12	\\
{Na \sc i}	&  \multicolumn{2}{c}{ }	&		&  \multicolumn{2}{c}{ }	&		&  \multicolumn{2}{c}{$-$0.44}	&	1	\\
\hline
\end{tabular}
\tablefoot{\tablefoottext{a}{Number of lines used per ion, the stated error is the rms error computed if two or more lines are available.}}
\end{table}

\subsection{Spectral fitting}\label{s-specfit}
F and G type MS companions of sdB stars can be easily resolved in the optical as they have similar brightnesses and distinct spectra. This allows one to disentangle such composite spectra from a single observation without knowing the radii or fluxes of the components. Such a binary decomposition was implemented in the {\sc XTgrid} \citep{Nemeth12} spectral fitting algorithm and was used to estimate the atmospheric parameters of the components in 29 composite spectra binaries.
{\sc XTgrid} employs the NLTE model atmosphere code {\sc TLUSTY} \citep{Hubeny95} for the subdwarf component and interpolated {\sc MILES} \citep{Cenarro07} template spectra for the MS companion. The binary spectrum is fitted with a linear combination of the two components. This method is independent from the SED and VWA analysis and was applied for low-resolution flux calibrated spectra obtained with the B\&C spectrograph, therefore it can be used to check the consistency of the different approaches.

\begin{figure*}[!t]
\centering
\includegraphics{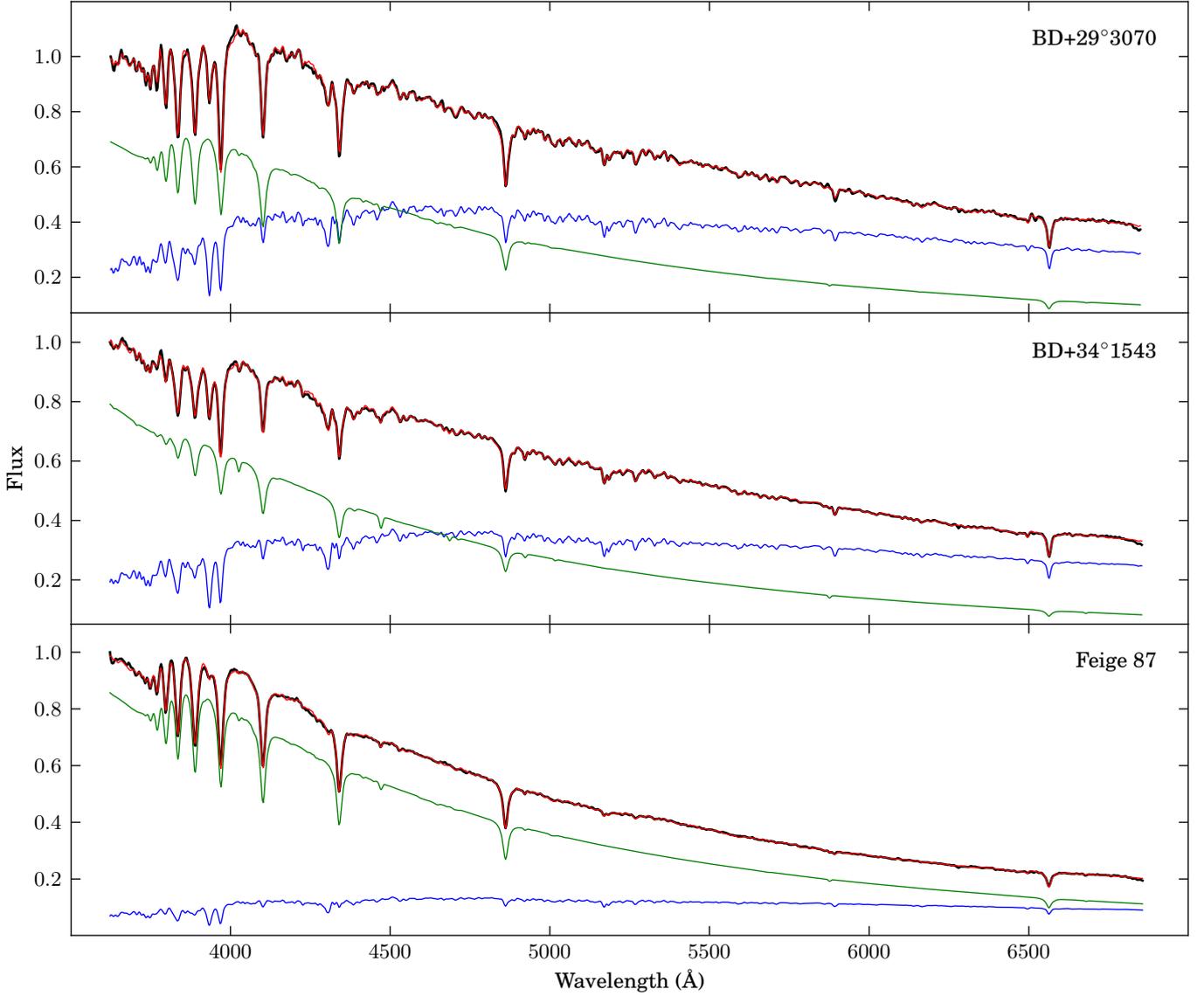}
\caption{The flux calibrated B\&C spectra observed with the Bok telescope (black line) of BD+29$^{\circ}$3070, BD+34$^{\circ}$1543 and Feige\,87 together with the best fitting binary atmosphere models from the {\sc XTgrid} code (red line). The decomposition of the model in the MS and sdB component is plotted in respectively blue and green.}\label{fig-greenspec}
\end{figure*}

Without preliminary assumptions on the spectral types {\sc XTgrid} confirmed the results of the SED and VWA analysis on BD$+$29$^\circ$3070 and BD$+$34$^\circ$1543 within error bars, but predicted a lower surface gravity of the components in Feige 87. Then, with the help of the radial velocity measurements, we could constrain the surface gravities of the companions which helped achieving a better consistency of the decomposition in all three cases. Our results are listed in Table \ref{tb-gspecresults} and the disentangled binary spectra are plotted in Fig. \ref{fig-greenspec}. We note that the lower contribution of the MS star in Feige 87 raises the uncertainties of our parameter determination.

%

\begin{table}
\caption{The atmospheric parameters obtained from the flux calibrated spectra of BD+29$^{\circ}$3070, BD+34$^{\circ}$1543 and Feige\,87.}
\label{tb-gspecresults}
\centering
\begin{tabular}{lrrr}
\hline\hline
\noalign{\smallskip}
Parameter & BD+29$^{\circ}$3070 & BD+34$^{\circ}$1543 & Feige\,87 \\\hline
\noalign{\smallskip}
E($B-V$) (mag)			&	0.013		&	0.002		&	0.000		\\
Flux ratio\tablefootmark{a}	&	1.353		&	1.337		&	2.252		\\
\noalign{\smallskip}
\multicolumn{4}{c}{MS component}\\
T$_{\rm{eff}}$ (K)		&	6026 $\pm$ 300	&	5715 $\pm$ 300	&	5675 $\pm$ 250	\\
$\log{g}$ (dex)			&	4.26 $\pm$ 0.30	&	4.11 $\pm$ 0.30	&	4.23 $\pm$ 0.35	\\
$[$Fe/H$]$			&	0.08 $\pm$ 0.25	&	-0.37$\pm$ 0.25	&	-0.39 $\pm$ 0.25\\
\noalign{\smallskip}
\multicolumn{4}{c}{sdB component}\\
T$_{\rm{eff}}$ (K)		&	25380 $\pm$ 990	&	36640 $\pm$ 810	&	27270 $\pm$ 500	\\
$\log{g}$ (dex)			&	5.54 $\pm$ 0.18	&	6.13 $\pm$ 0.16	&	5.47 $\pm$ 0.15	\\
He\tablefootmark{b}     &    -2.63$^{+0.48}_{-1.26}$&    -1.49 $\pm$ 0.13&    -2.56$^{+0.22}_{-0.50}$\\
C\tablefootmark{b}      &    -3.27 $>$             &    -4.64 $>$        &    -3.77 $>$             \\
N\tablefootmark{b}      &    -2.79 $>$             &    -3.28 $>$        &    -3.69 $>$             \\
O\tablefootmark{b}      &    -1.93 $>$             &    -3.72 $>$        &    -2.89 $>$             \\
\hline
\end{tabular}
\tablefoot{\tablefoottext{a}{Flux ratio F$_{\rm{sdB}}$ / F$_{\rm{MS}}$ in wavelength range 6720 - 6800 \AA}
\tablefoottext{b}{Abundance given as $\log(nX/nH)$}}
\end{table}

\section{Gravitational redshift}\label{s-gr}
If the difference between the surface gravity of both components in a binary system is substantial, this can give rise to a frequency shift in the emitted radiation. This effect is caused by a difference in gravitational redshift for the two stars. From general relativity one can derive that the gravitational redshift as function of the mass and surface gravity of the star \citep{Einstein16}:
\begin{equation}
 z_g = \frac{1}{c^2} \sqrt{G M g}. \label{e-gr}
\end{equation}
Where $z_g$ is the gravitational redshift, $c$ the speed of light, $G$ the gravitational constant, $M$ the mass, and $g$ the surface gravity. This $z_g$ will effectively change the apparent systemic velocity for the star. In a binary system the difference in surface gravity for both components will be visible as a difference in systemic velocity between the components. As $z_g$ is proportional to the square root of the surface gravity, this effect is only substantial when there is a large difference in $\log{g}$ between the components, as is the case for compact subdwarfs and main-sequence stars.

The difference in systemic velocity between the MS and sdB component measured in the radial velocity curves can be used to estimate the surface gravity of the sdB component (see also \citetalias{Vos12}). Using a canonical value of 0.47 $M_{\odot}$ for the sdB component and the mass ratio, the mass of the MS component can be calculated. Together with the surface gravity of the MS component derived from the SED fit and spectral analysis, the $z_g$ of the MS component can be calculated. The $z_g$ of the sdB component can then be derived by combining $z_{g,\rm{MS}}$ with the measured difference in systemic velocity, and can be converted to an estimated surface gravity for the sdB component.
A caveat here is that the wavelength of the He I $\lambda$ 5875 multiplet is only known with a presision of $\sim$ 0.01 \AA, corresponding to a systematic radial velocity uncertainty of up to $\sim$ 0.5 $\rm{km\,s}^{-1}$. The measured shift in systemic velocity could be partly due to this uncertainty, but since the observations are consistent with the predicted gravitational redshift we believe this to be the main contributor.

For BD+29$^{\circ}$3070, the system velocity of the sdB component is weakly constrained and does not yield a very constrained surface gravity for the sdB component. However, the system velocities of the other two systems are better constrained. The measured differences in system velocity for the three systems are:
\begin{equation}
 \gamma_{\rm{sdB}} - \gamma_{\rm{MS}} = \left\{
 \begin{array}{lr}
  0.73 \pm 1.46\ \rm{km\,s}^{-1} & (\rm{BD+29}^{\circ}3070) \\
  1.01 \pm 0.52\ \rm{km\,s}^{-1} & (\rm{BD+34}^{\circ}1543) \\
  1.34 \pm 0.51\ \rm{km\,s}^{-1} & (\rm{Feige\,87})
 \end{array}
 \right.
\end{equation}
Using the averaged values for the surface gravity of the MS components as given in Table \ref{tb-absolutedim}, the estimated surface gravities of the sdB components of these systems are:
\begin{equation}
\log(g)_{\rm{sdB}} = \left\{
 \begin{array}{lr}
  5.41^{+0.30}_{-0.62}\ \rm{dex} & (\rm{BD+29}^{\circ}3070)\\
  \noalign{\smallskip}
  5.45^{+0.14}_{-0.34}\ \rm{dex} & (\rm{BD+34}^{\circ}1543)\\
  \noalign{\smallskip}
  5.73^{+0.12}_{-0.37}\ \rm{dex} & (\rm{Feige\,87})
 \end{array}
 \right.
\end{equation}
The asymmetrical error on the surface gravity is calculated using a Monte Carlo simulation taking into account the error on the difference in system velocity and the errors on the parameters of the MS component. The resulting surface gravities correspond with the results from the other methods presented in this paper.

\section{Absolute parameters}\label{s-absolutepar}

\begin{table*}
\centering
\caption{Fundamental properties for both the main-sequence (MS) and subdwarf (sdB) component of BD+29$^{\circ}$3070, BD+34$^{\circ}$1543 and Feige\,87.}\label{tb-absolutedim}
\begin{tabular}{l@{\hskip 1cm}rr@{\hskip 1cm}rr@{\hskip 1cm}rr}
\hline\hline
\noalign{\smallskip}
	  & \multicolumn{2}{c}{BD+29$^{\circ}$3070} & \multicolumn{2}{c}{BD+34$^{\circ}$1543} & \multicolumn{2}{c}{Feige\,87}\\
Parameter & \multicolumn{1}{c}{MS} & \multicolumn{1}{c}{sdB} & \multicolumn{1}{c}{MS} & \multicolumn{1}{c}{sdB} & \multicolumn{1}{c}{MS} & \multicolumn{1}{c}{sdB} \\	
\hline
\noalign{\smallskip}
$P$ (d)			&	\multicolumn{2}{c}{1283 $\pm$ 63}    	&	\multicolumn{2}{c}{972 $\pm$ 2}    	&	\multicolumn{2}{c}{936 $\pm$ 2}    	\\
$T_0$			&	\multicolumn{2}{c}{2453877  $\pm$ 41}	&	\multicolumn{2}{c}{2451519  $\pm$ 11}	&	\multicolumn{2}{c}{2453259  $\pm$ 21}	\\
$e$			&	\multicolumn{2}{c}{0.15  $\pm$ 0.01} 	&	\multicolumn{2}{c}{0.16  $\pm$ 0.01} 	&	\multicolumn{2}{c}{0.11  $\pm$ 0.01} 	\\
$\omega$		&	\multicolumn{2}{c}{1.60  $\pm$ 0.22} 	&	\multicolumn{2}{c}{1.58  $\pm$ 0.07} 	&	\multicolumn{2}{c}{2.92  $\pm$ 0.15} 	\\
$\gamma$ (km s$^{-1}$)	&	\multicolumn{2}{c}{$-$57.58 $\pm$ 0.36}	&	\multicolumn{2}{c}{32.10 $\pm$ 0.06}	&	\multicolumn{2}{c}{32.98 $\pm$ 0.08}	\\
$q$			&	\multicolumn{2}{c}{0.39 $\pm$ 0.01}  	&	\multicolumn{2}{c}{0.57 $\pm$ 0.01}  	&	\multicolumn{2}{c}{0.55 $\pm$ 0.01}  	\\
$a$ ($R_{\odot}$)	&	\multicolumn{2}{c}{586 $\pm$ 12}  	&	\multicolumn{2}{c}{447 $\pm$ 4}  	&	\multicolumn{2}{c}{442 $\pm$ 3}  	\\
$i$ $(^o)$		&	\multicolumn{2}{c}{81 $\pm$ 5}		&	\multicolumn{2}{c}{43 $\pm$ 1}		&	\multicolumn{2}{c}{75 $\pm$ 2}		\\
$E(B-V)$		&	\multicolumn{2}{c}{0.010 $\pm$ 0.030}	&	\multicolumn{2}{c}{0.005 $\pm$ 0.030}	&	\multicolumn{2}{c}{0.006 $\pm$ 0.030}	\\
$d$ (pc)		&	\multicolumn{2}{c}{238 $\pm$ 25}	&	\multicolumn{2}{c}{207 $\pm$ 30}	&	\multicolumn{2}{c}{383 $\pm$ 40}	\\
$[$Fe/H$]$		&	\multicolumn{2}{c}{0.05 $\pm$ 0.16}	&	\multicolumn{2}{c}{$-$0.26 $\pm$ 0.08}	&	\multicolumn{2}{c}{-0.50 $\pm$ 0.18}	\\
\noalign{\smallskip}	
$K$ (km s$^{-1})$   	&	6.53 $\pm$ 0.30	&	16.6 $\pm$ 0.6 	&	5.91 $\pm$ 0.07	&	10.31 $\pm$ 0.15&	8.19 $\pm$ 0.11	&	15.01 $\pm$ 0.21\\
$M$ ($M_{\odot})$	&	1.19 $\pm$ 0.09	&	0.47 $\pm$ 0.05	&	0.82 $\pm$ 0.07	&	0.47 $\pm$ 0.05	&	0.86 $\pm$ 0.07	&	0.47 $\pm$ 0.05	\\
$\log{g}$ (cgs)		&	4.32 $\pm$ 0.50	&	5.56 $\pm$ 0.44	&	4.18 $\pm$ 0.40	&	5.84 $\pm$ 0.35	&	4.36 $\pm$ 0.42	&	5.54 $\pm$ 0.34	\\
$R$ ($R_{\odot})$	&	1.25 $\pm$ 0.97	&	0.19 $\pm$ 0.10	&	1.21 $\pm$ 0.48	&	0.14 $\pm$ 0.06	&	1.02 $\pm$ 0.50	&	0.19 $\pm$ 0.07	\\
$T_{\textrm{eff}}$ (K)	&	6180 $\pm$ 420	&	25900 $\pm$ 3000&	6100 $\pm$ 300	&	36600 $\pm$ 3000&	5980 $\pm$ 325	&	27300 $\pm$ 2700\\
$L$ ($L_{\odot}$)	&	2.06 $\pm$ 0.25	&	14.3 $\pm$ 3.0	&	1.83 $\pm$ 0.25	&	30.13 $\pm$ 4.0	&	1.19 $\pm$ 0.20	&	18.04 $\pm$ 3.0	\\
$V_0 $ (mag)		&	10.91$\pm$ 0.10	&	11.47$\pm$ 0.15	&	10.72$\pm$ 0.10	&	11.11$\pm$ 0.15	&	12.54$\pm$ 0.10	&	12.36$\pm$ 0.15	\\
$M_V$   (mag) 		&	4.02 $\pm$ 0.10	&	4.58 $\pm$ 0.15	&	4.14 $\pm$ 0.10	&	4.54 $\pm$ 0.15	&	4.62 $\pm$ 0.10	&	4.44 $\pm$ 0.15	\\
\hline
\end{tabular}
\tablefoot{The sdB mass is the canonical value.}
\end{table*}

Combining the results of the SED fit, the spectral analysis, gravitational redshift and the orbital parameters derived from the radial velocity curves, the absolute dimensions of BD+29$^{\circ}$3070, BD+34$^{\circ}$1543 and Feige\,87 can be determined. With the assumed canonical sdB mass of $M_{\rm{sdB}}$ = 0.47 $M_{\odot}$, the inclination of the systems can be derived from the reduced mass determined in Sect. \ref{s-orbitalparam}. This inclination can be used to calculate the semi-major axis of the systems. 
The atmospheric parameters ($T_{\rm{eff}}$, $\log{g}$) determined with the four different methods (SED fitting, derived from the iron lines, spectral fitting and from the gravitational-redshift) correspond well within their errors. The final values for these parameters are the averages of the three methods weighted by their errors. The radii of the components is calculated from the mass and surface gravity. The system velocity is set to the system velocity of the MS component as the sdB component is subjected to significant gravitational redshift as discussed in Section \ref{s-gr}.

The luminosity of both components can be calculated using $L = 4\pi \sigma R^2 T^4$. The apparent V magnitudes are obtained directly from the SED fitting procedure, while the absolute magnitude can be obtained by integrating the best fit model SEDs over the Johnson V band, and scaling the resulting flux to a distance of 10 pc. The distance to the system is then calculated from $\log{d} = (m_V - M_V + 5)/5$. For BD+34$^{\circ}$1543 the distance obtained in this way corresponds within errors with the distance obtained from the Hipparcos parallax 236 $\pm$ 80 pc. The absolute dimensions of all three systems are summarized in Table \ref{tb-absolutedim}.

The proper motions of BD+29$^{\circ}$3070, BD+34$^{\circ}$1543 and Feige\,87 as measured by \citet{Vanleeuwen07} are:
\begin{eqnarray}
  (\mu_{\alpha}, \mu_{\delta}) &=& (-6.29, 23.92) \pm (1.00, 1.55)\ \rm{mas\ yr^{-1}}\\
  (\mu_{\alpha}, \mu_{\delta}) &=& (34.46, -61.08) \pm (2.28, 1.42)\ \rm{mas\ yr^{-1}}\\
  (\mu_{\alpha}, \mu_{\delta}) &=& (14.12, -65.51) \pm (1.49, 1.82)\ \rm{mas\ yr^{-1}}
\end{eqnarray}
Using the method of \citet{Johnson87}, these numbers together with the measured value of $\gamma$, can be used to compute the galactic space velocity vector with respect to the local standard of rest from \citet{Dehnen98}. Resulting in
\begin{eqnarray}
(U,V,W)_{\rm{LSR}} &=& (-43.0,-27.1, -6.8) \pm (3.1, 1.5, 1.8)\ \rm{km\ s^{-1}}\\
(U,V,W)_{\rm{LSR}} &=& (-13.0,-69.9, 28.1) \pm (0.9, 4.8, 2.3)\ \rm{km\ s^{-1}}\\
(U,V,W)_{\rm{LSR}} &=& ( 95.1,-32.5, 94.1) \pm (13.9, 8.5, 9.0)\ \rm{km\ s^{-1}}
\end{eqnarray}
for respectively BD+29$^{\circ}$3070, BD+34$^{\circ}$1543 and Feige\,87. U is defined as positive towards the galactic center. Following the selection criteria of \citet{Reddy06}, all systems are bound to the galaxy, and belong to the thin or thick disk population. 

\section{Eccentric orbits}
The three systems presented here are part of a sample consisting of six sdB + MS binaries, of which four systems have been analysed up till now. The first system PG\,1104$+$243 was described in \citetalias{Vos12}, and is the only one from our sample with a circular orbit ($P$ = 753 d, $e$ $<$ 0.002). The two remaining systems (Balloon\,82800003 and BD$-$7$^{\circ}$5977) have longer periods that have not yet been covered completely at all phases, but preliminary results were presented in \citet{Vos13} demonstrating that both systems are significantly eccentric. Furthermore \citep{Barlow13} analysed PG\,1449$+$653 ($P$ = 909 d, $e$ = 0.11) and PG\,1701+359 ($P$ = 734 d, circular), and \citep{Deca12} published results of PG\,1018$+$243 which has a possible eccentric orbit ($P$ = 759 d, $e$ $\approx$ 0.2). At the time of this publication, there are nine long period sdB + MS systems know of which six have a significant eccentric orbit, one might have an eccentric orbit and two have a circular orbit.

Current theory predicts only circular orbits for these long period sdB binaries formed through the stable RLOF channel. During the red giant phase of the sdB progenitor the tidal forces between the sdB progenitor and its companion should circularize the orbit very efficiently \citep{Zahn77}, and the further evolution provides very few possibilities to re-introduce eccentricity to the orbit. \citet{Deca12} proposed the hierarchial triple merger scenario of \citet{Clausen11} for the possibly eccentric sdB + K system PG\,1018-047, where the K-type companion would not have been involved in the evolution of the sdB component. However, such a scenario seems too unlikely to be observed frequently.

Another possible explanation for these eccentric systems would be that tidal circularization during the RGB and the RLOF phase is not as efficient as currently assumed. In highly eccentric systems the mass transfer through RLOF and mass loss due to stellar winds will not be constant over the orbit. \citet{Bonacic08} studied this effect in binary systems with an AGB star, and found that enhanced mass loss from the AGB star at orbital phases closer to periastron can work efficiently against the tidal circularization of the orbit. Further theoretical studies could investigate if this eccentricity enhancement mechanism can work in in the progenitors of sdB + MS binaries as well.

A last interesting observation of this small sample is that the two circular systems and PG\,1018$+$243 with a possible circular orbit, have the shortest orbital periods. Although the sample is too small to state strong conclusions, this might support the above claim that the circularization process is not as efficient as expected and doesn't fully circularize the longer period systems.

\section{Conclusions}\label{s-summary}
Using both literature photometry and observed spectra, detailed astrophysical parameters of BD+29$^{\circ}$3070, BD+34$^{\circ}$1543 and Feige\,87 have been established. The long time-base spectroscopic observations made it possible to determine accurate periods, and to solve the orbit of both the MS and sdB component in all three systems. The atmospheric parameters were determined with three different techniques: based on the spectral energy distribution, using iron lines in the disentangled MS spectra, and from low-resolution flux calibrated spectra. Furthermore, the measured gravitational-redshift could be used to derive the surface gravity of the sdB components. The results obtained with these different methods agree within their errors.

The sdB components of all three systems are consistent with a canonical post-core-helium-flash model with a mass around 0.47 $M_{\odot}$. However, the orbits of these systems are clearly eccentric, as opposed to what the current theory predicts. The periods found here correspond with the updated version of the BPS studies of \citet{Chen13}, although a significant degree of atmospheric RLOF must be included in order to reach the $\sim$1300\,d period of BD+29$^{\circ}$3070. Furthermore, \citet{Chen13} finds a metallicity -- orbital period relation, where the orbital period will decrease with decreasing metallicity. The observed periods of BD+29$^{\circ}$3070, BD+34$^{\circ}$1543 and Feige\,87 (1283\,d, 972\,d, 936\,d) for metallicities (solar, about half solar, about 1/3$^{\rm{rd}}$ solar) follow the metallicity relation,  but significantly exceed the predictions of \citet{Chen13}. The same conclusion can be reached for PG\,1104$+$243 (P = 753\,d, $\sim$1/3$^{\rm{rd}}$ solar, \citetalias{Vos12}). 
We assume that the metallicity measured for the companion is comparable to the initial Z for the system, which ignores any contamination during RLOF, and can therefore be considered as an upper limit to the initial Z. All four systems call for the inclusion of atmospheric RLOF, also at lower metallicities. While the sample is too limited yet to make conclusions with respect to the period distribution, the long periods found in this paper indicates that an additional mechanism such as atmospheric RLOF is required for all the systems, not just those with periods $>$1100 d, as suggested by \citet{Chen13}.

BD+29$^{\circ}$3070, BD+34$^{\circ}$1543 and Feige\,87 are part of an ongoing long-term observing program of sdB + MS binaries with HERMES at Mercator. Currently most of the systems accessible with the 1.2m Mercator + HERMES telescope-spectrograph combination have been analysed, but orbital solutions of more systems are needed in order to test the predicted orbital distributions. Our results call for the study of eccentricity pumping mechanisms (e.g. \citealt{Bonacic08}) also in the framework of sdB binaries.

\begin{acknowledgements}
  Based on observations made with the Mercator Telescope, operated on the island of La Palma by the Flemish Community, at the Spanish Observatorio del Roque de los Muchachos of the Instituto de Astrofísica de Canarias.
  Based on observations obtained with the HERMES spectrograph, which is supported by the Fund for Scientific Research of Flanders (FWO), Belgium , the Research Council of K.U.Leuven, Belgium, the Fonds National Recherches Scientific (FNRS), Belgium, the Royal Observatory of Belgium, the Observatoire de Genève, Switzerland and the Thüringer Landessternwarte Tautenburg, Germany.
  The research leading to these results has received funding from the European Research Council under the European Community's Seventh Framework Programme (FP7/2007--2013)/ERC grant agreement N$^{\underline{\mathrm o}}$\,227224 ({\sc prosperity}), as well as from the Research Council of K.U.Leuven grant agreements GOA/2008/04 and GOA/2013/012, the German Aerospace Center (DLR) under grant agreement 05OR0806 and the Deutsche Forschungsgemeinschaft under grant agreement WE1312/41-1.
  The following Internet-based resources were used in research for this paper: the NASA Astrophysics Data System; the SIMBAD database and the VizieR service operated by CDS, Strasbourg, France; the ar$\chi$ive scientific paper preprint service operated by Cornell University.
  This publication makes use of data products from the Two Micron All Sky Survey, which is a joint project of the University of Massachusetts and the Infrared Processing and Analysis Center/California Institute of Technology, funded by the National Aeronautics and Space Administration and the National Science Foundation.
\end{acknowledgements}

\bibliographystyle{aa}
\bibliography{references}

\end{document}